\documentclass[twocolumn]{aastex701}
\usepackage[english]{babel}
\usepackage[utf8]{inputenc}
\usepackage[colorinlistoftodos, color=green!40, prependcaption]{todonotes}
\usepackage{subfigure}
\linenumbers
\usepackage[shortlabels]{enumitem}

\usepackage{appendix}

\begin{document}

\title{Very high-energy gamma-ray and neutrino emission from hadronic interaction in compact binary millisecond pulsars}

% Discussing the detection possibility of very high-energy gamma-ray emission from hadronic interaction in compact binary millisecond pulsars

\correspondingauthor{V. Vecchiotti}
%\email{vittoria.vecchiotti@ntnu.no}
\email{vittoria.vecchiotti@inaf.it}

\author{V. Vecchiotti}
\affiliation{NTNU, Department of Physics, NO-7491 Trondheim, Norway}
%\email{vittoria.vecchiotti@ntnu.no}
\email{vittoria.vecchiotti@inaf.it}

\author{M. Linares}
\affiliation{NTNU, Department of Physics, NO-7491 Trondheim, Norway}
\affiliation{Departament de F{\'i}sica, EEBE, Universitat Polit{\`e}cnica de Catalunya, Av. Eduard Maristany 16, E-08019 Barcelona, Spain.}
\email{manuel.linares@ntnu.no}

%\author{A. K. Harding}
%\affiliation{Theoretical Division, Los Alamos National Laboratory, Los Alamos, NM 87545, USA}

\begin{abstract}
Blackwidow and redback systems are millisecond pulsars in compact orbits with ultra-light and low-mass companions, respectively, collectively known as ``spider pulsars". 
In such systems, an intrabinary shock can form between the pulsar and the companion winds, serving as a site for particle acceleration and associated non-thermal emission. 
Assuming that protons can be extracted from the neutron star surface and accelerated at the intrabinary shock and/or within the pulsar wind, we model the very high-energy gamma-ray and neutrino emissions ($0.1-10^3$~TeV) produced through interactions with the companion wind and the companion star.
We first calculate the high-energy emissions using an optimistic combination of parameters to maximize the gamma-ray and neutrino fluxes. 
We find that, for energetic spider pulsars with a spin-down power $\gtrsim 10^{35}\rm erg\, s^{-1}$ and a magnetic field of $\sim 10^{3}\, \rm G$ in the companion region, the gamma-ray emission could be detectable as point sources by CTA and LHAASO, while the neutrino emission could be detectable by the future TRIDENT detector.
Finally, we build a synthetic population of these systems, compute the cumulative neutrino flux expected from spider pulsars, and compare it with the Galactic neutrino diffuse emission measured by IceCube. We find that, under realistic assumptions on the fraction of the spin-down power converted into protons, the contribution of spiders to the diffuse Galactic neutrino flux is negligible.
\end{abstract} 


%\maketitle

\section{Introduction}

In recent years, an increasing number of millisecond pulsars (MSPs) have been detected in the GeV energy range by Fermi-LAT \citep{Smith_2023}. 
% Spider systems
Among these systems, a significant subset consists of MSPs in compact binary systems with orbital periods $P_b\lesssim1 $ day, where the companion stars are not fully degenerate \citep{Nedreaas2024}.
Due to their small orbital separations, typically just a few Solar radii, the pulsar’s relativistic wind can strongly irradiate and ablate the companion star. 
These systems, known as spiders, are classified into two categories based on the companion star's mass: redbacks (RBs) and blackwidows (BWs). RB companions typically have masses $M_{\rm c}\gtrsim 0.1 M_{\odot}$, while BW companions are even lighter, with  $M_{\rm c} < 0.1 M_{\odot}$. 

% The intrabinary shock
Many spiders exhibit a non-thermal, orbitally modulated emission component in the X-ray band that peaks at either inferior or superior conjunction \citep{Huang_2012, Bogdanov_2005, Bogdanov_2011, Bogdanov_2021,Wadiasingh_2017,Wadiasingh_2018}.
The X-ray emission is often associated with the presence of an intrabinary pulsar wind termination shock (IBS), where the ram pressure of the pulsar wind is counterbalanced by that of the companion's wind or magnetosphere \citep{harding1990}.
The IBS is considered an efficient site for particle acceleration \citep{harding1990, Arons1993, Cortes:2022tzh, Cortes:2024ibe,Cortes:2025wjv}. 
While the exact acceleration mechanism remains uncertain, it may involve either diffusive shock acceleration (e.g., \citealt{harding1990,vanderMerwe2020ApJ}) or magnetic reconnection (e.g., \citealt{vanderMerwe2020ApJ,Cortes:2022tzh, Cortes:2024ibe,Cortes:2025wjv}).
In either case, the observed X-ray flux is thought to result from synchrotron emission produced by relativistic electron-positron pairs accelerated at the IBS and Doppler boosted along the shock tangent (e.g., \citealt{Wadiasingh_2017,Wadiasingh_2018,Romani2016,Sanchez_2017,vanderMerwe2020ApJ,Kandel_2019,Kandel2021}).
Additionally, these electron-positron pairs responsible for X-ray emission may also interact via inverse Compton scattering with the radiation field of the companion star, producing orbitally modulated TeV gamma-rays that are potentially detectable by current and future gamma-ray facilities \citep{vanderMerwe2020ApJ}. 
However, no spider systems have been detected in the TeV gamma-ray band to date.

% Orbital modulation in the GeV energy range
Instead, in the GeV energy range,  gamma-ray emission from pulsar binaries is predominantly driven by pulsed magnetospheric radiation originating from the pulsar itself and is generally expected to be orbitally constant (apart from the pronounced dips observed in the light curves of eclipsing systems; \citealt{Corbet:2022xbf,Clark:2023owb}).
Moreover, in some RB systems orbitally modulated GeV signals have been detected with the Fermi-LAT in a few systems (e.g., \citealt{Ng_2018, An_2018, Clark:2020hbv}), suggesting a potential IBS origin.
However, in some RB systems, the GeV gamma-ray emission appears to be anti-correlated with the X-ray emission.  
This anti-correlation suggests that alternative or additional mechanisms might contribute to the observed GeV modulation.
In order to explain the GeV signal, various scenarios have been proposed \citep{An:2020qsf,Sim:2024kxi}.
For instance, the orbitally modulated GeV gamma rays could originate from synchrotron emission produced by highly energetic leptons that, after being accelerated at the IBS, may penetrate the companion star's region and interact with its magnetic field, producing the observed signal \citep{Sim:2024kxi}.

% This work
In this work, we explore the possibility that, in addition to electron-positron pairs, protons are also extracted from the neutron star's surface. 
The possibility of baryons extracted from the neutron star surface was invoked in \cite{Hoshino1992ApJ} and \cite{Gallant1994ApJ}, in order to explain particle acceleration at the
pulsar wind termination shock and was further explored in later studies (e.g., \citealt{Lemoine:2014ala,Kotera:2015pya, Philippov:2017ikm, Guepin:2019fjb}).
The gamma-ray and neutrino emission produced as a result of the interaction of these accelerated hadrons with a target material has been investigated in various studies, e.g., in the case of the Crab Nebula (see \citealt{Amato2003A&A}).

In spider systems, hadrons can be accelerated either at the IBS \citep{harding1990} or, as in the pulsar case, within the pulsar magnetosphere and then get advected with the pulsar wind (PW) \citep{Kotera:2015pya}. 
Here, we consider both acceleration scenarios and examine the possibility that these hadrons penetrate the companion region, where they interact with the companion's wind (CW) and the companion star (CS). Such interactions produce gamma rays and neutrinos.

We investigate whether spiders can generate gamma rays and neutrinos at levels detectable by current and future gamma-ray and neutrino observatories. 
Furthermore, we assess whether the cumulative neutrino flux from a synthetic Galactic spider population could contribute significantly to the Galactic neutrino flux measured by IceCube \citep{IceCubeScience}.

%The paper is organized as follows. In Sec.~\ref{sec:Model} and Sec.~\ref{sec:transport}, we introduce the different acceleration scenarios and outline the transport equation. In Sec.~\ref{Sec:Gamma and nu}, we describe the methodology for calculating gamma-ray and neutrino production.  In Sec.~\ref{sec: population}, we detail the generation of a synthetic population of spiders. In Sec.~\ref{sec:Result}, we present and discuss our results. Finally, we summarize our conclusions in Sec.~\ref{Sec:Conclusion}.

The paper is organized as follows. In Sec.~\ref{sec:Model} and Sec.~\ref{sec:interaction}, we introduce the different acceleration scenarios and the interaction sites.
In Sec.~\ref{sec:transport}, we outline the transport equation. In Sec.~\ref{Sec:Gamma and nu}, we describe the methodology for calculating gamma-ray and neutrino production.  In Sec.~\ref{sec: population}, we detail the generation of a synthetic population of spiders. In Sec.~\ref{sec:Result}, we present and discuss our results. Finally, we summarize our conclusions in Sec.~\ref{Sec:Conclusion}.

\section{Proton acceleration scenarios}
\label{sec:Model}
In this section, we model the high-energy gamma rays and neutrinos being produced by protons extracted from the neutron star surface. We consider two scenarios:
\begin{enumerate} % a), b)
\item[1)] Scenario 1: PW. The protons are all accelerated in the pulsar magnetosphere (e.g., \citealt{Guepin:2019fjb}) and then advected with the PW. Consequently, the injected spectrum is a delta function with energy equal to the Lorentz factor of the wind (e.g., \citealt{Kotera:2015pya}).
\item[2)] Scenario 2: IBS. The protons are accelerated at the IBS via magnetic reconnection \citep{Cortes:2022tzh, Cortes:2024ibe,Cortes:2025wjv, vanderMerwe2020ApJ}. In this case, the injected spectrum is a power law with an exponential cut-off, where the spectral index $\alpha$ lies within the range $1.5$ to $2.5$, and the wind magnetization defines the cut-off energy.
\end{enumerate}
%{\bf Explain why the p gamma is not possible} \\
In both cases, we assume that all protons can penetrate the companion region due to their high energy and consequently large Larmor radius.
%{\bf This is true only if the magnetic field is in the downstream region of the shock is sufficiently small}
%
Once inside the companion region (the downstream region of the shock), protons are expected to diffuse due to the turbulent magnetic field.  
In this environment, they can interact either with the CW or with the CS, leading to the production of gamma rays and neutrinos.

In the following, we briefly describe the two acceleration scenarios considered in this work and the target material.

\begin{figure*}
\begin{center}
\subfigure[]
{\includegraphics[width=0.45\textwidth]{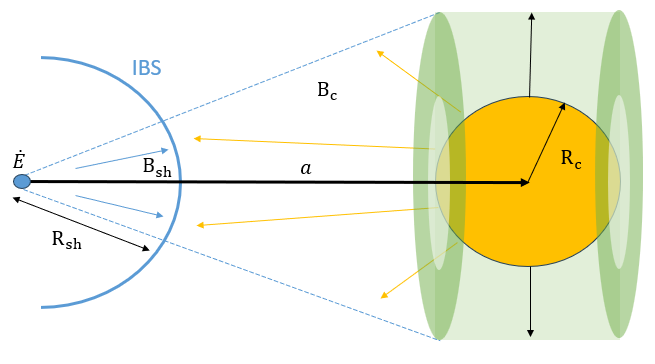}}
\subfigure[]
{\includegraphics[width=0.45\textwidth]{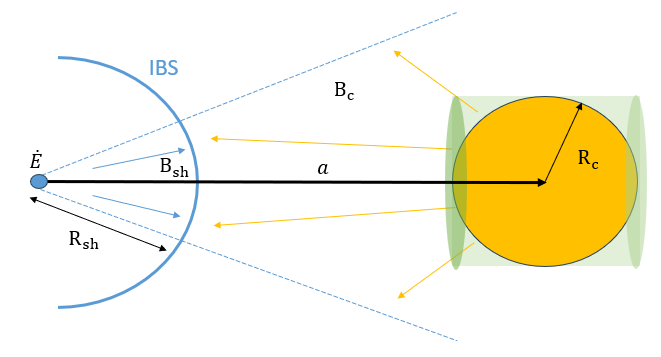}}
\caption{\small\em  
Sketch of a spider system. The millisecond pulsar is represented as a blue sphere on the left (not to scale), while the companion star is shown as a yellow sphere on the right. 
The pulsar (companion) wind is indicated with blue (yellow) arrows. The interaction between the winds forms the IBS, depicted as a blue semicircle.
Panel (a): The green hollow cylinder around the companion represents the interaction region of protons with the CW.
Panel (b): The green cylinder around the companion represents the interaction region of protons with the CS.
\label{fig: Sketch}}
\end{center}
\end{figure*}

\subsection{Scenario 1: protons from the PW}
The mechanism that accelerates particles in the wind to the terminal Lorentz factor remains a matter of debate. In this work, we assume that the wind energy is primarily dominated by particle kinetic energy, rather than by Poynting flux (see, e.g., \citealt{Kotera:2015pya}). This is equivalent to assuming that a large fraction, $\eta \sim 1$, of the pulsar spin-down energy is converted into particle energy.

In standard scenarios, particles are typically accelerated within the pulsar magnetosphere, either close to the neutron star surface and/or near the light cylinder, or in the current sheet (see e.g., \citealt{Philippov:2020jxu}).
% acceleration in the current sheet?
However, inside the light cylinder, in the presence of a background photon field, ions may experience significant curvature losses \citep{Kotera:2015pya}.
Following \cite{Kotera:2015pya}, we assume that, regardless of the maximum energy achieved in the pulsar magnetosphere, once the particles reach the wind region, they are advected with it at the wind’s Lorentz factor $\Gamma_{\rm w}$ that we will calculate later on in the text (see Sec.~\ref{sec:transport}).
As a result, in this scenario, all protons will be injected with energy equal to the Lorentz factor of the wind.

\subsection{Scenario 2: protons accelerated at the IBS}
The second scenario that we consider in this work is that hadrons are accelerated at the IBS.
In general, collisionless shocks are considered efficient sites for diffusive shock acceleration, producing particles with a power-law spectrum and an exponential cutoff.
The spectral index $\alpha$ is typically larger than $2$, while the cut-off energy depends on shock properties such as radius, magnetic field, and compression ratio. 
However, the IBS is a relativistic shock that is expected to be strongly magnetized and quasi-perpendicular, making diffusive shock acceleration particularly challenging and inefficient (see e.g., \cite{Sironi:2015oza} for a review).
Indeed, the non-thermal X-ray spectrum of most spider systems exhibits a relatively flat photon index $\Gamma_{\rm x}=1-2$ (e.g,. \citealt{Linares:2014aaa, Gentile2014ApJ}). 
This observation suggests a harder electron energy spectrum ($\alpha<2$), which is difficult to explain within the framework of diffusive shock acceleration.
The most likely acceleration mechanism taking place at the IBS is magnetic reconnection \citep{Cortes:2022tzh, Cortes:2024ibe, Cortes:2025wjv}.
The pulsar wind consists of toroidal stripes of alternating magnetic field polarity, separated by current sheets of hot plasma \citep{Bogovalov:1999vg, Petri:2007tz}. 
When oppositely directed fields are compressed at the shock, they can undergo annihilation through shock-driven reconnection \citep{Cortes:2022tzh, Cortes:2024ibe, Cortes:2025wjv}.
For example, \cite{Cortes:2024ibe} have shown that if the pulsar spin axis is nearly aligned with the orbital angular momentum, the magnetic energy of the relativistic pulsar wind is efficiently converted into particle energy at the IBS. 
%The highest energy particles accelerated by reconnection can stream ahead of the shock and be further accelerated by the upstream motional electric field. In the downstream, further energization is governed by stochastic interactions with the plasmoids/magnetic islands generated by reconnection.

In this paper, to reflect these uncertainties on the acceleration mechanism, we explore two possible values for the proton spectral index: $\alpha=1.5$ and $\alpha=2.5$. 
We calculate the cut-off energy assuming that protons are accelerated by magnetic reconnection (see Sec.~\ref{sec:transport} \& Eq.~\ref{eq: Ecut}). 
However, we note that a small fraction of particles
%, after being accelerated by reconnection, 
may undergo further acceleration to higher energies through Fermi-like processes \citep{Cortes:2024ibe}.

\section{Interaction sites}
\label{sec:interaction}

\begin{deluxetable}{l|cc}
\tablenum{1}
\tablecaption{\em 
We report the typical parameters for RB (second column) and BW (third column) systems. In order, we display the companion mass $M_{\rm c}$, the orbital period $P_{\rm b}$, the companion radius $R_{\rm c}$, the orbital separation $a$,  the shock radius $R_{\rm sh}$, the wind number density $n_{\rm w}$ for $\dot{E}=10^{33}-10^{35} \rm erg \, s^{-1}$, and the companion number density $n_{\rm c}$.
\label{tab: standard parameters}}
\tablewidth{0pt}
\tablehead{
\nocolhead{.} & \colhead{RB} & \colhead{BW}}
%\decimalcolnumbers
\startdata
%%%%%%%%%%%%%%%%%%%%%%%%%%%%%%%%%%%%%%%%%%%
$M_{\rm c}$ & $0.1 M_{\odot}$ & $0.01 M_{\odot}$ \\
 \hline
%%%%%%%%%%%%%%%%%%%%%%%%%%%%%%%%%%%%%%%%%%%
$P_{\rm b}$ & $8.2\, \rm h$ & $4.9\, \rm h$ \\
 \hline
%%%%%%%%%%%%%%%%%%%%%%%%%%%%%%%%%%%%%%%%%%%
$R_{\rm c}$ & $3.1\times 10^{10}\,\rm cm$ & $1.0\times 10^{10}\,\rm cm$ \\
 \hline
%%%%%%%%%%%%%%%%%%%%%%%%%%%%%%%%%%%%%%%%%%%
$a$ & $1.7\times 10^{11}\,\rm cm$ & $1.2\times 10^{11}\,\rm cm$  \\
 \hline
%%%%%%%%%%%%%%%%%%%%%%%%%%%%%%%%%%%%%%%%%%%
$R_{\rm sh}$ & $7.7\times 10^{10}\,\rm cm$ & $6.6\times 10^{10}\,\rm cm$  \\
 \hline
%%%%%%%%%%%%%%%%%%%%%%%%%%%%%%%%%%%%%%%%%%%
$n_{\rm w}$ & $3.1\times10^{9-11}\, \rm cm^{-3}$ & $4.0\times10^{10-12}\, \rm cm^{-3}$ \\
 \hline
%%%%%%%%%%%%%%%%%%%%%%%%%%%%%%%%%%%%%%%%%%%
$n_{\rm c}$ & $9.8\times10^{23}\, \rm cm^{-3}$ & $2.5\times10^{24}\, \rm cm^{-3}$ \\
\enddata
\end{deluxetable}

% characteristic of the spiders (target):
In order to estimate the hadronic gamma-ray and neutrino fluxes from spiders, 
we model two interaction sites that the accelerated protons (Sec.~\ref{sec:Model}) will reach: the CW and the CS.
First, we collect, estimate, and list the main parameters of spider systems that are relevant to our calculation.
In particular, in the case of RBs (BWs), the companion mass $\rm M_{\rm c}$ is approximately $0.1 \, \rm M_{\odot}$ ($0.01 \,\rm M_{\odot}$).
As for the neutron star mass, we assume $1.7 \, \rm M_{\odot}$ \citep{Strader:2018qbi}.
In this work, we adopt for the orbital periods of RBs and BWs the average values obtained from a sample of known sources, as listed in Tables \ref{tab:Red Backs} and \ref{tab: Black widows} \citep{Nedreaas2024}.
We obtain orbital periods of $8.2\, \rm h$ for RBs and $4.9\,\rm h$ for BWs.
Additionally, we assume that the CS is filling its Roche Lobe.
With the above information, we can calculate the radius of the CS being $R_{\rm c}\sim 3.1\times 10^{10}\, \rm cm$ for RBs and $R_{\rm c}\sim 1.0\times 10^{10}\, \rm cm$ for BWs.
Using the above values, we calculate the average number density of the CS, $n_{\rm c}\sim 9.8\times 10^{23}\, \rm cm^{-3}$ and $n_{\rm c}\sim 2.5 \times 10^{24}\, \rm cm^{-3}$ for RBs and BWs, respectively.
We can also estimate the average number density of the CW in the following way. The mass loss rate in the CW can be parameterized as
\begin{equation}
\dot{M}_{w}=f_w \dot{E}\left(\frac{R_{\rm c}}{4 G M_{\rm c}}\right)\left(\frac{R_{\rm c}}{a}\right)^{2},
\label{Eq: Mdot}
\end{equation}
where $f_w=0.1$ is the fraction of the irradiating luminosity intercepted by the companion that goes into launching a thermal wind \citep{10.1093/mnras/254.1.19P} and $a$ is the orbital separation between the pulsar and the companion.
The number density of the wind is then:
\begin{equation}
n_{\rm w}(r, \dot{E})=\frac{\dot{M}_{w}}{4 \pi r^2 v_{\rm w} m_p },
\end{equation}
where $r$ is the distance from the center of the companion, $v_{\rm w}$ is the escape velocity from the companion and $m_p$ is the proton mass.
In order to estimate an upper limit on the wind number density, we calculate $n_{\rm w}$ at the companion surface $r=R_2$ for a range of spin-down power $10^{33}-10^{35} \, \rm erg \, s^{-1}$.
The chosen spin-down power range reflects the variability observed in Tables \ref{tab:Red Backs} and \ref{tab: Black widows}.
 Using Eq.~\ref{Eq: Mdot}, we calculate the shock radius according to Eq.~\ref{Eq:shockRadius} in Appendix \ref{IBS}.
The last relevant quantity is the magnetic field in the companion region, $\rm B_{\rm c}$.
\cite{Sanchez:2017qcl} suggest that the companion may have a strong magnetic field; however, this quantity is currently poorly constrained. 
In this work, we remain agnostic and instead explore a broad range of possible values, $\rm B_{\rm c}=0.1-10^{3}\,\rm G$.
All the above-calculated quantities are summarized in Tab.~\ref{tab: standard parameters}.

In the case of protons interacting with the CS, the gamma rays produced are immediately absorbed by the dense companion. The path length of a $1$ TeV photon in an environment with number density $\sim 10^{24} \, \rm cm^{-3}$ is $\sim 40 \rm \, cm$.
Therefore we find that observable gamma rays are produced exclusively through the interaction of accelerated protons with the CW. 
In contrast, the majority of the neutrino signal arises from interactions with the CS, due to the significantly higher target proton density.
From this point forward, we will calculate gamma rays from proton interactions with the CW and neutrinos from proton interactions with the CS.

\section{Transport equation}
\label{sec:transport}
In order to calculate the proton spectrum at the interaction site, we consider a simple leaky box model. 
%For both scenarios, we consider a simplified setup in which the particles are injected at the boundary of the box and diffuse due to the turbulent magnetic field inside it.
For both scenarios, we consider a simplified setup in which the particles are injected isotropically from the neutron star. Only those that intersect the interaction regions (displayed in green in Figure \ref{fig: Sketch}) are injected into the box and subsequently diffuse away from the pulsar due to the turbulent magnetic field within the box.
%In case of interaction with the CS, we consider a cylinder centered on the companion, with the normal to the base of the cylinder approximately aligned with the direction of the pulsar wind  (see panel (b) of Fig.~\ref{fig: Sketch}). 
In case of interaction with the CS, we center the interaction region on the companion.  For simplicity, we represent this region as a cylinder, with the normal to the base approximately aligned with the direction of the pulsar wind  (see panel (b) of Fig.~\ref{fig: Sketch}). 
We set the radius of the cylinder equal to $R_{\rm c}$ and the height equal to $2 R_{\rm c}$.
%For interaction with the CW, we consider a hollow cylinder surrounding the companion, aligned as the previous one.  Specifically, we set $R_{\rm inner}=R_{\rm c}$ and $R_{\rm outer}=2 R_{\rm c}$, and the height equal to $2 R_{\rm c}$ (see panel (a) of Fig.~\ref{fig: Sketch}).
For interaction with the CW, the interaction occurs in the densest part of the wind, which is located close to the CS.
Thus, we model this as a hollow cylinder surrounding the companion, aligned as the previous one.  Specifically, we set $R_{\rm inner}=R_{\rm c}$ and $R_{\rm outer}=2 R_{\rm c}$, and the height equal to $2 R_{\rm c}$ (see panel (a) of Fig.~\ref{fig: Sketch}).
We emphasize that Fig.~\ref{fig: Sketch} is provided to illustrate where the interaction takes place. However, the geometrical details do not directly enter our calculation, which depends only on the intersected surface and the length along the axis connecting the pulsar to the CS, along which propagation occurs.

We solve the following transport equation for the number of protons with energy $E$ at a given time t, $N(E,t)$:
\begin{equation}
%\nonumber
\frac{\partial N(E, t)}{\partial t}= - \frac{N(E, t)}{\tau_{\rm pp}(E)} - \frac{N(E, t)}{\tau_{\rm esc}(E)}+ Q(E, t)
%\frac{\partial N(E, t)}{\partial t} = - \frac{\partial}{\partial E} \left[ b(E, t) N(E, t) \right] - \frac{N(E, t)}{\tau(E, t)} + Q(E, t)
\label{Eq:transport}
\end{equation}
where $Q(E, t)$ is the injected proton spectrum, $\tau_{\rm pp}$ is the loss timescale due $pp$ interaction with the CW and/or the CS and $\tau_{\rm esc}$ is the escape timescale.
In particular, $\tau_{\rm pp}$ is defined as
\begin{equation}
\tau_{\rm pp}(E) 
= (n_{i} c \sigma_{pp})^{-1},
\end{equation}
where $n_{i}$ is the number density of the target $i$ (c: companion or w: companion wind), $c$ is the speed of light and $\sigma_{pp}$ is the total cross-section of the $pp$ interaction as parameterized by \cite{Kelner:2006tc}.
While the escape timescale is
\begin{equation}
\tau_{\rm esc}= \max(\tau_{\rm diff},\tau_{\rm b}),
\label{eq:escape timescale}
\end{equation}
where $\tau_{\rm diff}$ and $\tau_{\rm b}$ represent the diffusion and ballistic timescales, respectively.
We take the maximum of the two to ensure that particles do not diffuse faster than the speed of light.
% The only energy loss in this environment is due to $pp$ interaction with the wind and the companion star.
%
The diffusion timescale, assuming Bohm diffusion, is given by
\begin{equation}
\tau_{\rm diff}(E) = \frac{3 R^2}{c} \frac{eB_{\rm c}}{E},
\label{eq: diffusion timescale}
\end{equation}
where $B_{\rm c}$ is the magnetic field in the companion region (i.e., the downstream region of the IBS), $e$ is the electron charge and $R$ is the size of the box parallel to the direction of propagation of the pulsar wind. We assume that particles are escaping from the base of the cylinders.
Thus, $R=2 R_{\rm c}$ for interactions with both the PW and the CS. 
The ballistic timescale is
\begin{equation}
\tau_{\rm b} = \frac{R}{c}.
\end{equation}

% solution of the transport equation
In order to solve Eq.~\ref{Eq:transport}, we assume stationarity. This is a reasonable assumption since in the case of millisecond pulsars, the spin-down power remains stable over timescales of $\sim \rm Gyr$ \citep{2004hpa..book.....L}.
From Eq.~\ref{Eq:transport}, we get:
\begin{equation}
N(E)=Q(E)\tau_{\rm eff}(E)
%0 = - \frac{ N(E, t)}{\tau_{\rm pp}(E, t)} - \frac{N(E, t)}{\tau(E, t)} + Q(E, t)
\label{Eq:solution transport}
\end{equation}
where $\tau_{\rm eff}= (\tau_{\rm pp}^{-1}+\tau_{\rm esc}^{-1})^{-1}$.

%\subsection{Injection}
%The acceleration mechanism at the intrabinary shock is still under debate...
%
For the two considered scenarios, we have two possibilities for the injected proton spectrum: 1) all protons injected with the Lorentz factor of the PW, and 2) protons injected with a power-law spectrum (IBS).

In {\bf scenario 1: PW}, the injected spectrum can be expressed as
\begin{equation}
Q(E,t) = \dot{N}_{\rm GJ}(t)\delta(E-E_{\rm max}(t)),     
\end{equation}
where $\dot{N}_{\rm GJ}(t)$ is the Goldreich-Julian rate defined in Eq.~\ref{eq:GJ} and  $E_{\rm max}(t)$ is the energy of the protons flowing with the wind:
\begin{equation}
E_{\rm max}(t)=\Gamma_{\rm w} m_{\rm p} c^{2},
\label{eq:Emax}
\end{equation}
where $\Gamma_{\rm w}$ is the wind Lorentz factor. 
%In order to define this last quantity, we follow the approach of \cite{Kotera:2015pya}.
%
In our model, a large fraction $\eta\sim1$ of the pulsar spin-down power $\dot{E}$ is converted into kinetic luminosity.
%In particular, we assume that the luminosity gets distributed among pairs and protons in the following way \citep{Kotera:2015pya}:
%\begin{equation}
%\eta\dot{E}= 2 k \dot{N}_{\rm GJ} m_{\rm e} c^2 \Gamma_{\rm w}\left(1+\frac{m_{\rm p}}{2 k m_{\rm e}}\right)
%\label{eq:energy}
%\end{equation}
%  where $k$ is the pair multiplicity, namely the number of pairs produced for each electron extracted from the neutron star surface.
%Using Eq.~\ref{eq:energy}, we can then write the wind Lorentz factor as:
In particular, we assume that the luminosity gets distributed among pairs and protons and hence we can write the wind Lorentz factor as \citep{Kotera:2015pya}:
\begin{equation}
\Gamma_{\rm w}=\frac{\eta \dot{E}}{2 k \dot{N}_{\rm GJ} m_{\rm e} c^2 \left( 1+\frac{m_{\rm p}}{2 k m_{\rm e}}\right)}\\
%\nonumber
%&=& \left(\frac{8 \pi^4}{3 c^4}\right)^{\frac{1}{2}}\left(\frac{ \eta e}{2 m_{\rm e}}\right) k^{-1}B P^{-2}R^{3}\left(1+\frac{t}{\tau_{\rm sd}}\right)^{-1}\left(1+\frac{m_{\rm p}}{2 k m_{\rm e}}\right)^{-1}
\label{eq:GammaWind}
\end{equation}
where $k$ is the pair multiplicity, namely the number of pairs produced for each electron extracted from the neutron star surface.
See Sec.~\ref{sec: multiplicity} for the resulting ranges of $\Gamma_{\rm w}$ and $E_{\rm max}$ in spiders.

In {\bf scenario 2: IBS}, the injected spectrum is defined as
\begin{equation}
Q(E,t) = Q_{0}(t) E^{-\alpha}exp \left({-\frac{E}{E_{\rm cut}}}\right), 
\end{equation}
where $\alpha=1.5-2.5$, $E_{\rm cut}$ is the cut-off energy,
\begin{equation}
Q_{0}(t) = \frac{\eta_{\rm p} \dot{E}(t)}{\int_{E_{\rm min}}^{\infty}dE \, E Q(E,t)}  
\label{Eq:Q0}
\end{equation}
is the normalization factor with $\eta_{\rm p}$ being the fraction of the spin-down power that goes into protons. For $E_{\rm min}$ we assume the value $1$ GeV (the rest energy of protons).
The average post-shock Lorentz factor achievable by particles via magnetic reconnection is $\Gamma\sim \Gamma_{\rm w} \sigma$ assuming efficient dissipation in the striped pulsar wind.
Hence, the cut-off energy is approximately:
\begin{equation}
E_{\rm cut}\simeq\Gamma m_{\rm p}c^2=\Gamma_{\rm w} \sigma   m_{\rm p}c^2
\label{eq: Ecut}
\end{equation}
where $\sigma$ is the magnetization parameter (i.e., the ratio of Poynting to kinetic energy flux).
%and $\Gamma_{\rm w}$ is, once again, the wind Lorentz factor.
%
The magnetization in the upstream region of the IBS can be estimated from:
\begin{equation}
\sigma=\frac{B_{\rm sh}^2}{8 \pi \Gamma_{\rm w} (k m_{e} c^2+m_p c^2)\rho_{GJ}(R_{\rm sh})}
\label{eq: magnetization}
\end{equation}
where $B_{sh}$ is the magnetic field in the upstream region of the IBS shock, $R_{\rm sh}$ is the radius of the IBS, $\rho_{GJ}(R_{\rm sh})$ is the Goldreich-Julian density at the IBS location, $k$ is the pair multiplicity, $m_{e}$ and $m_p$ are the electron and proton mass, respectively.
The magnetic field is dipolar inside the pulsar light cylinder $R_{\rm LC}=c P_0/(2\pi)$ with $P_0$ initial period, and toroidal in the wind, hence:
\begin{equation}
B_{\rm sh}=B_{0}\left(\frac{R_{\rm NS}}{R_{\rm LC}}\right)^{3}\left(\frac{R_{\rm sh}}{R_{\rm LC}}\right)^{-1}
\label{eq: B shock}
\end{equation}
where $B_{0}$ is the magnetic field at the neutron star surface of a spider system, $R_{\rm NS}=10$ km is the radius of the neutron star.
The Goldreich-Julian density is calculated accordingly to:
\begin{equation}
\rho_{GJ}(R_{\rm sh})=\frac{B_{\rm LC}}{P_0 c e}\left(\frac{R_{\rm LC}}{R_{{\rm sh}}}\right)^2
\label{eq: GJ density}
\end{equation}
where $B_{\rm LC}= B_0 (R_{\rm NS}/R_{\rm LC})^3$ is the magnetic field at the pulsar light cylinder.

See Sec.~\ref{sec: magnetization} for the resulting ranges of $\sigma$ and $E_{\rm cut}$ in spiders.

\begin{figure*}
\begin{center}
\subfigure[]{\includegraphics[width=0.45\textwidth]{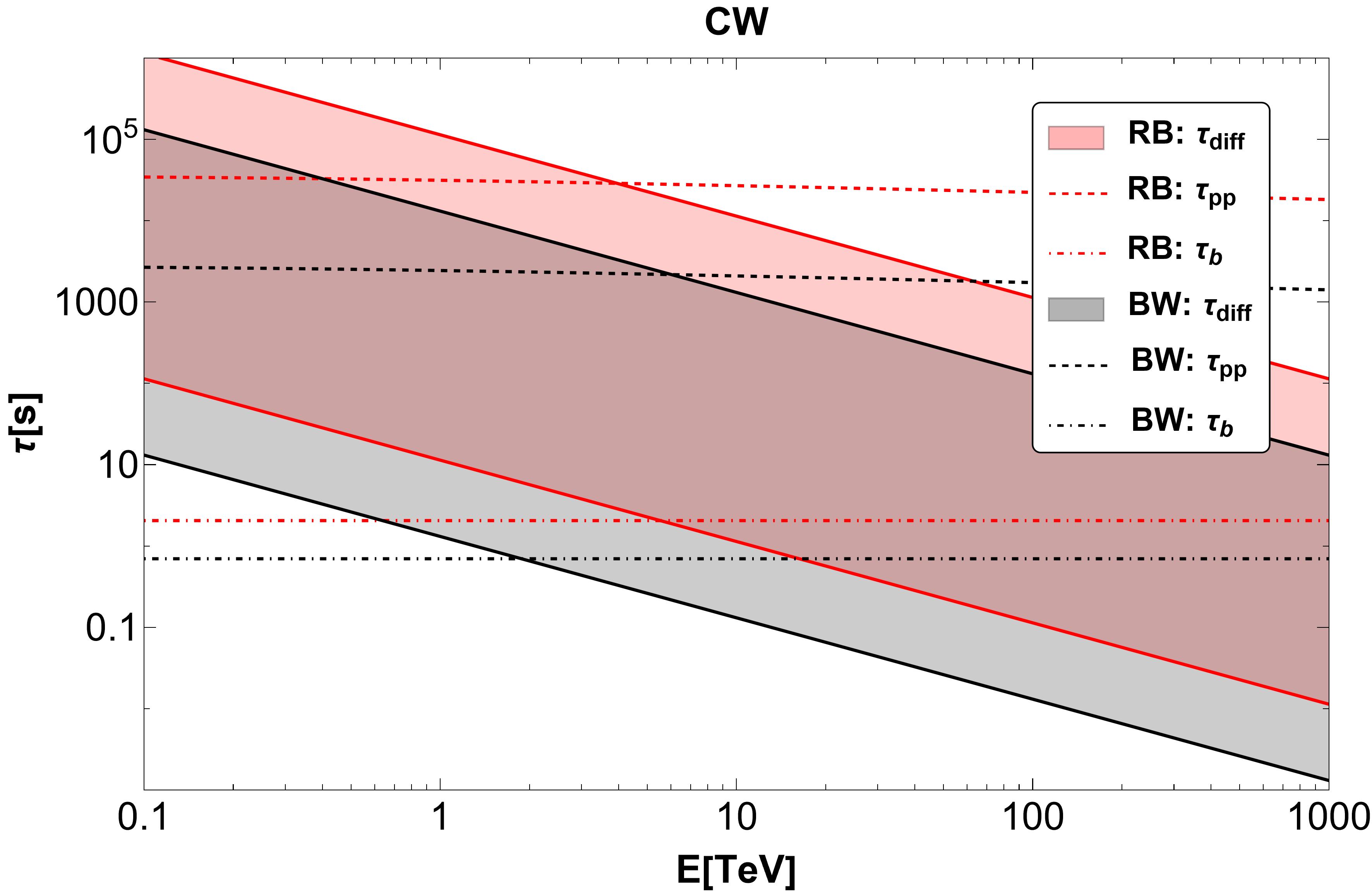}}
\subfigure[]{\includegraphics[width=0.45\textwidth]{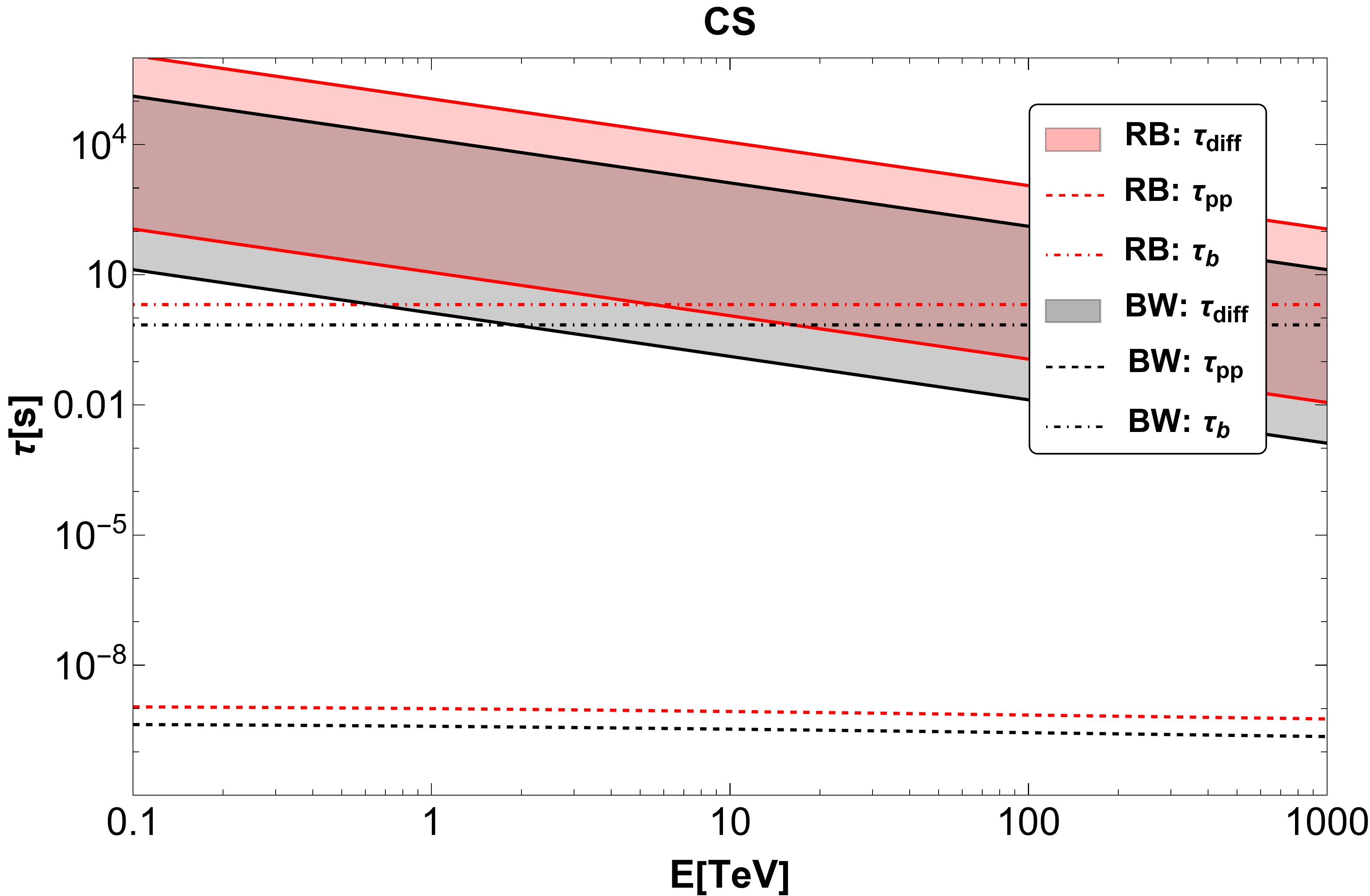}}
\caption{\small\em Timescales for the interaction of protons with the CW (panel (a)) and with the CS (panel (b)).
The diffusion, pp, and ballistic timescales are displayed with colored-shadowed bands, and dashed, dash-dotted lines, respectively.
Timescales for typical RBs (BWs) are displayed in red (black).
The uncertainty in $\tau_{\rm diff}$ is due to the uncertainty on the magnetic field in the companion region.
The lower part of the band is obtained assuming $B_{\rm c}=0.1$ G, while the upper part is obtained for $B_{\rm c}=10^3$ G.
}.
\label{fig:timescale}
\end{center}
\end{figure*} 

\begin{figure}
%\begin{center}
\includegraphics[width=0.45\textwidth]{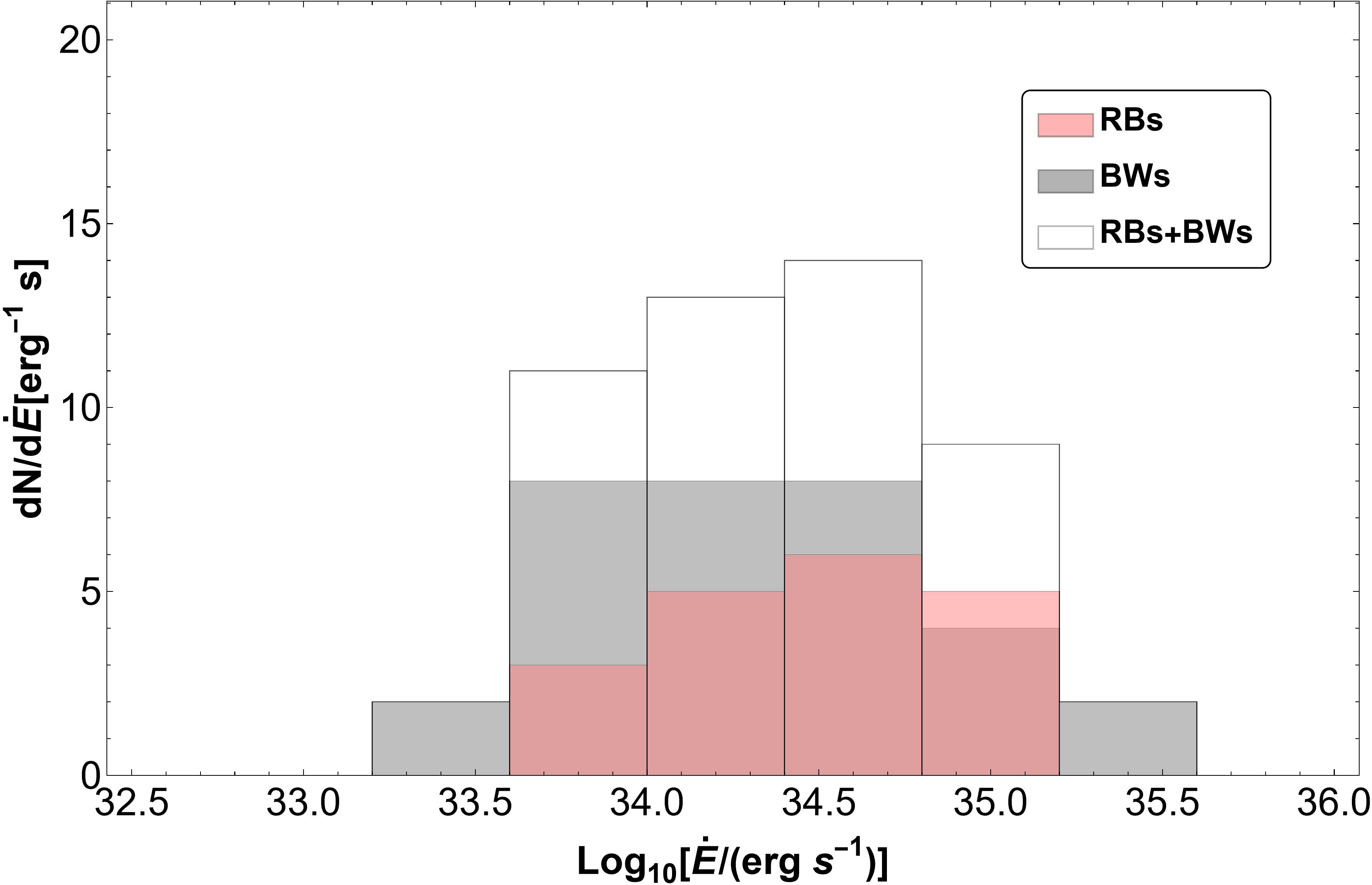}
\caption{\small\em  
Histogram of the spin-down power for RBs (red), BWs (black), and RBs plus BWs (white) in Tab.\ref{tab:Red Backs} and Tab.\ref{tab: Black widows} with known spin-down power.
}
\label{fig:histogram}
%\end{center}
\end{figure} 

\subsection{Diffusion vs Advection}
\label{sec: caveat}
In order for the particles that have crossed the IBS to reach the interaction region (depicted in green in Fig.~\ref{fig: Sketch}), diffusion must be faster than advection within the CW.
In other words, the diffusion timescale, $\tau_{\rm diff}(E) = 3 d^2 eB_{\rm c}/(cE)$, must be shorter than the advection timescale, $\tau_{\rm adv}=v_{\rm w}/d$, where $d=a-R_{\rm sh}-R_{\rm c}\sim \, 10^{10}$ cm represents the distance between the IBS and the interaction region.
For typical parameters of RBs and BWs (see Tab. \ref {tab: standard parameters}) we have $v_{\rm w} \sim 10^7 \rm cm \,s^{-1}$.
This comparison sets an upper limit on the magnetic field strength in the companion region, $B_{\rm c}$.
Ensuring that all protons reach the interaction region is especially crucial for neutrino production, which primarily arises from proton-CS interactions.
In contrast, this constraint is more relaxed for gamma rays, as the CW, though less dense, also occupies the region outside the green hollow cylinder in panel (a) of Fig.~\ref{fig: Sketch}, allowing gamma-ray production in this intermediate region as well.

By comparing the diffusion and advection timescales, we find that protons with $E>50$ TeV can always manage to reach the CS for the full range of $B_{\rm c}=0.1-10^3$ G considered in this work.
However, protons with $E\sim 10$ TeV can reach the CS only if $B_{\rm c}\lesssim 170$ G.
To ensure efficient neutrino production across the full TeV range, we therefore require $B_{\rm c}\le 170$ G.
As noted above, this condition is not as strict for gamma rays. In this work, we retain the full range of $B_{\rm c}$, since our goal is to identify the parameter ranges that maximize the flux on Earth. 
%For gamma rays, this maximum typically corresponds to the highest values of $B_{\rm c}$.

\subsection{Timescales}
\label{sec: timescales}
In Fig.~\ref{fig:timescale}, we show the timescales as a function of energy, for proton interaction with the CW (panel (a)) and with the CS (panel (b)).
In the CW, the escape timescale (Eq.~\ref{eq:escape timescale}) is the shortest and plays the major role while in the CS, the most relevant timescale is the one related to the $pp$ interaction.
This is because the companion is very dense, hence, the  $pp$ interaction is very fast. 
In this case (see panel (b)), the uncertainties on the magnetic field don't affect the final results.
As we note in Sec.~\ref{sec: caveat}, it is sufficient that $B_{\rm c}\le 170$ G to allow protons to reach the CS.
On the other hand, if we focus on panel (a), we can notice that this uncertainty plays a significant role.
The intensity of the magnetic field determines the energy $E_{\rm cross}$ at which $\tau_{\rm diff}=\tau_{\rm b}$, marking the transition from diffusive to ballistic propagation.
In particular, if the magnetic field is low ($B=0.1$ G, lower part of the band in Fig.~\ref{fig:timescale}), particles with $\sim 2$ TeV for BWs  ($\sim 5$ TeV for RBs) can already escape the box ballistically.  Instead, increasing the magnetic field up to $10^3$ G shifts this energy up to $>10^3$ TeV for both BWs and RBs.
Additionally, the magnetic field determines the normalization of the proton flux.
This is because the intensity of the magnetic field affects how long the particles stay trapped in the box before escaping.

In conclusion, the predicted gamma rays resulting from proton interactions with the CW are subject to significant uncertainties due to the unknown value of the companion's magnetic field.
In contrast, neutrinos generated from proton interactions with the CS remain unaffected by this parameter.
In the following, we show the effect of variation of $B_{\rm c}$ on the predicted gamma-ray signal, considering the range $B_{\rm c}=0.1-10^{3}$ G.

%In order to show the effect of variation of $B_{\rm c}$ on the predicted gamma-ray signal, we consider the range $B_{\rm c}=0.1-10^{3}$ G. {\color{magenta} reference to B fields estimates in spiders} 
%Another important ingredient is the magnetic field in the companion region $B_{\rm c}$, we fix it to an average value of $10$ Gauss. However, other values can be considered. 

\subsection{Pair multiplicity and fraction of the spin-down power converted into protons}
\label{sec: multiplicity}
% description of the pair multiplicity
Another uncertain parameter is the pair multiplicity $k$.
In particular, in our model for scenario 1: PW, we consider two possibilities for this value: one based on \cite{2011ApJHarding} and a higher value, closer to that typically assumed for pulsars. 
In \cite{2011ApJHarding}, the rate of electron and positron pairs in millisecond pulsars was calculated as:
\begin{equation}
\dot{N}_{\rm pair}= 3.1\times 10^{34}\, s^{-1} (\dot{E}_{35})^{0.68},\, \epsilon=0.6 
\label{eq: multiplicity}
\end{equation}
where the rate is increased by the presence of an offset polar cap with offset parameter $\epsilon \neq 0$ in the pulsar magnetosphere.
Using the above formulas, we calculate the pair multiplicity, $k=\dot{N}_{\rm pair}/\dot{N}_{\rm GJ}$, and find that $k=100$ for $\dot{E}=10^{33}\, \rm erg\, s^{-1}$ and $k=300$ for $\dot{E}=10^{35}\, \rm erg\, s^{-1}$. 
As $\dot{E}$ and the number of pairs increase, the energy available for protons decreases, and vice versa. With these values of $k$, we estimate that a fraction $\eta_{\rm p}\sim 0.9-0.8$ of the spin-down power is converted into protons, which is quite large. 
In this regard, we also present results for $k=10^{4}$, a higher value for the pair multiplicity, which gives $\eta_{\rm p}\sim 0.08$.

In the case of scenario 1: PW, the pair multiplicity affects the wind Lorentz factor $\Gamma_{\rm w}$ and the maximum proton energy $E_{\rm max}$ defined in Eq.~\ref{eq:GammaWind} and Eq.~\ref{eq:Emax}, respectively. 
In particular, assuming the multiplicity given by Eq.~\ref{eq: multiplicity}, we obtain a wind Lorentz factor of $\sim 5\times10^{4}$ ($\sim 4.5\times10^{5}$)
for a millisecond pulsar with spin-down power $\dot{E}=10^{33}\, \rm erg\, s^{-1}$ ($\dot{E}=10^{35}\, \rm erg\, s^{-1}$), which corresponds to a maximum proton energy of $\sim 50$ TeV ($\sim 400$ TeV).
If we instead consider $k=10^4$, we get $\Gamma_{\rm w}\sim4.9\times 10^3$ ($\Gamma_{\rm w}\sim4.9\times 10^4$) for a millisecond pulsar with spin-down power $\dot{E}=10^{33}\, \rm erg\, s^{-1}$ ($\dot{E}=10^{35}\, \rm erg\, s^{-1}$), which corresponds to $E_{\rm max}=4.5$ TeV ($E_{\rm max}=45$ TeV).

In scenario 1: PW, we treat the multiplicity $k$ as a free parameter, while in scenario 2: IBS, it is more suitable to consider $\eta_{\rm p}$ directly as a free parameter. This is because $\eta_{\rm p}$ enters the calculation as a pure normalization factor (see Eq.~\ref{Eq:Q0}) simply scaling our flux predictions up or down.

\subsection{Magnetization}
\label{sec: magnetization}
To assess the efficiency of magnetic reconnection in accelerating particles in spider systems, we estimate the magnetization parameter $\sigma$ in the upstream region of the IBS, which is directly related to the cutoff energy (see Eq.~\ref{eq: Ecut}).
By combining Eq.~\ref{eq: GJ density}, Eq.~\ref{eq: B shock} and Eq.~\ref{eq: magnetization} and plugging in typical values for the shock radius $R_{\rm sh}\sim 10^{11}$ cm, the magnetic field at the neutron star surface $B_{0}\sim  10^8$ G and the initial spin period $P_0\sim 10^{-3}$ s, we obtain magnetization values in the range $\sigma=3.2-31.7$ for $k=100$ and $\Gamma_{\rm w}=10^4-10^5$, and $\sigma=0.3-2.9$ for $k=10^4$ and the same range of the wind Lorentz factor.
These parameters yield average post-shock Lorentz factors of $\Gamma\sim 3\times 10^5$ for $k=100$ and $\Gamma\sim 3\times 10^4$ for $k=10^4$.
Assuming protons, the corresponding average particle energies are $\sim 300$ TeV and $\sim 30$ TeV for $\Gamma_{\rm}=3\times10^5$ and $\Gamma_{\rm}=3\times10^4$, respectively.
In this work, we adopt $E_{\rm cut}=300$ TeV, the maximum predicted value, as our standard reference case, as we aim to consider the most optimistic scenario for gamma-ray and neutrino detection.
%  
%The above estimate of the cutoff energy is physically motivated; however, 
It is important to note that $E_{\rm cut}$ remains somewhat uncertain, as it depends on the acceleration mechanism and poorly constrained parameters such as $B_{\rm sh}$.

The estimated range of $\sigma$ is consistent with what is assumed in the simulation by \cite{Cortes:2024ibe}, who adopt $\sigma=10$. 
%In our case, the magnetization is lower due to the inclusion of protons (see Eq.~\ref{eq: magnetization}). In the absence of protons, we obtain $\sigma=6.7-67$ for $k=100$.

%
The post-shock magnetic field value is uncertain, and therefore we assume a broad range $0.1-10^3$ G.
This leads to a downstream magnetization $\sigma\ll10^{-3}$ that is compatible with what is required by observations (see \citealt{vanderMerwe2020ApJ}). In this framework, nearly all the magnetic energy is converted into particle kinetic energy; hence, we can fairly assume $\eta\sim1$. 

\begin{figure*}
\begin{center}
\subfigure[]{\includegraphics[width=0.45\textwidth]{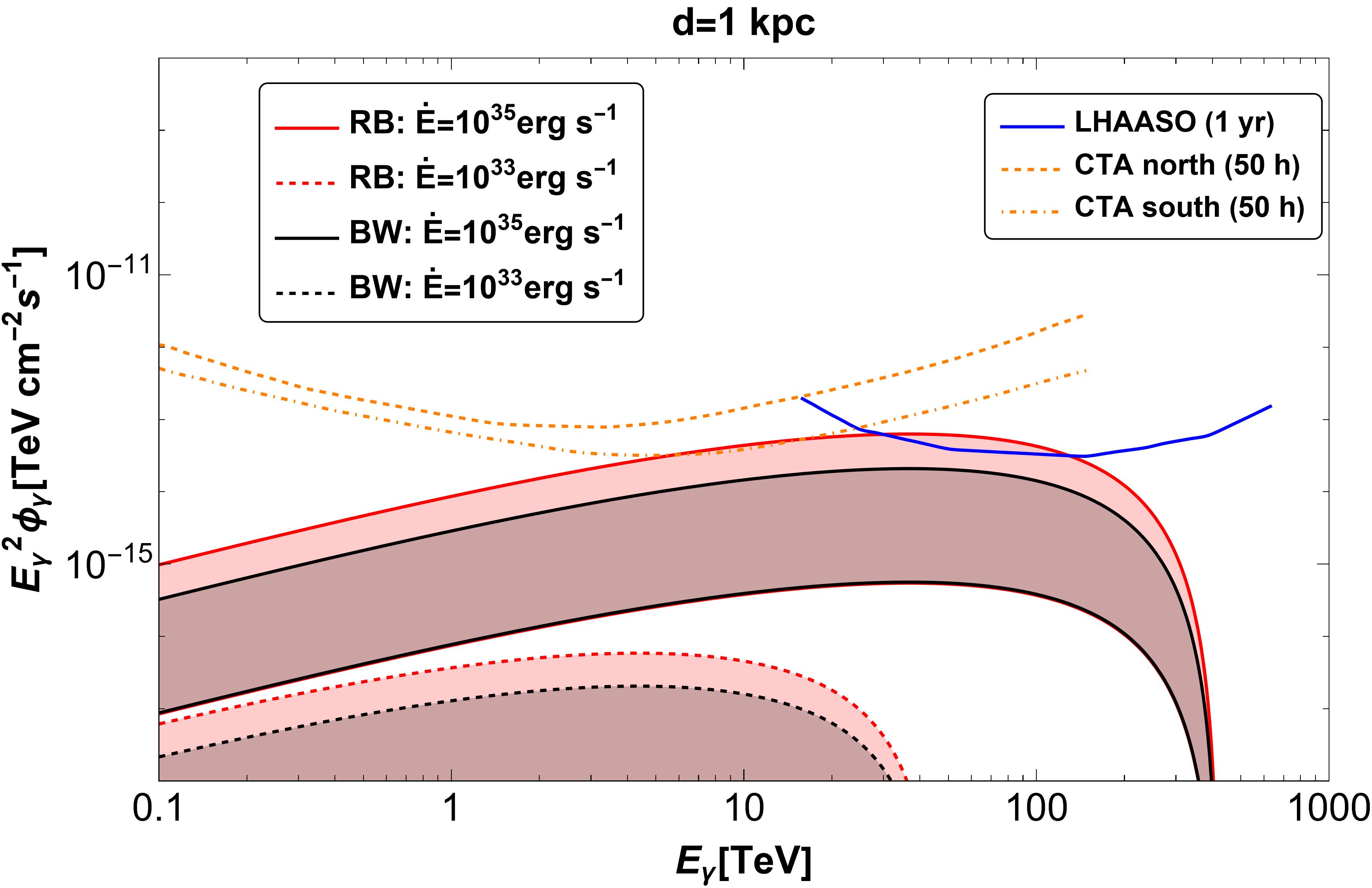}}
\subfigure[]{\includegraphics[width=0.45\textwidth]{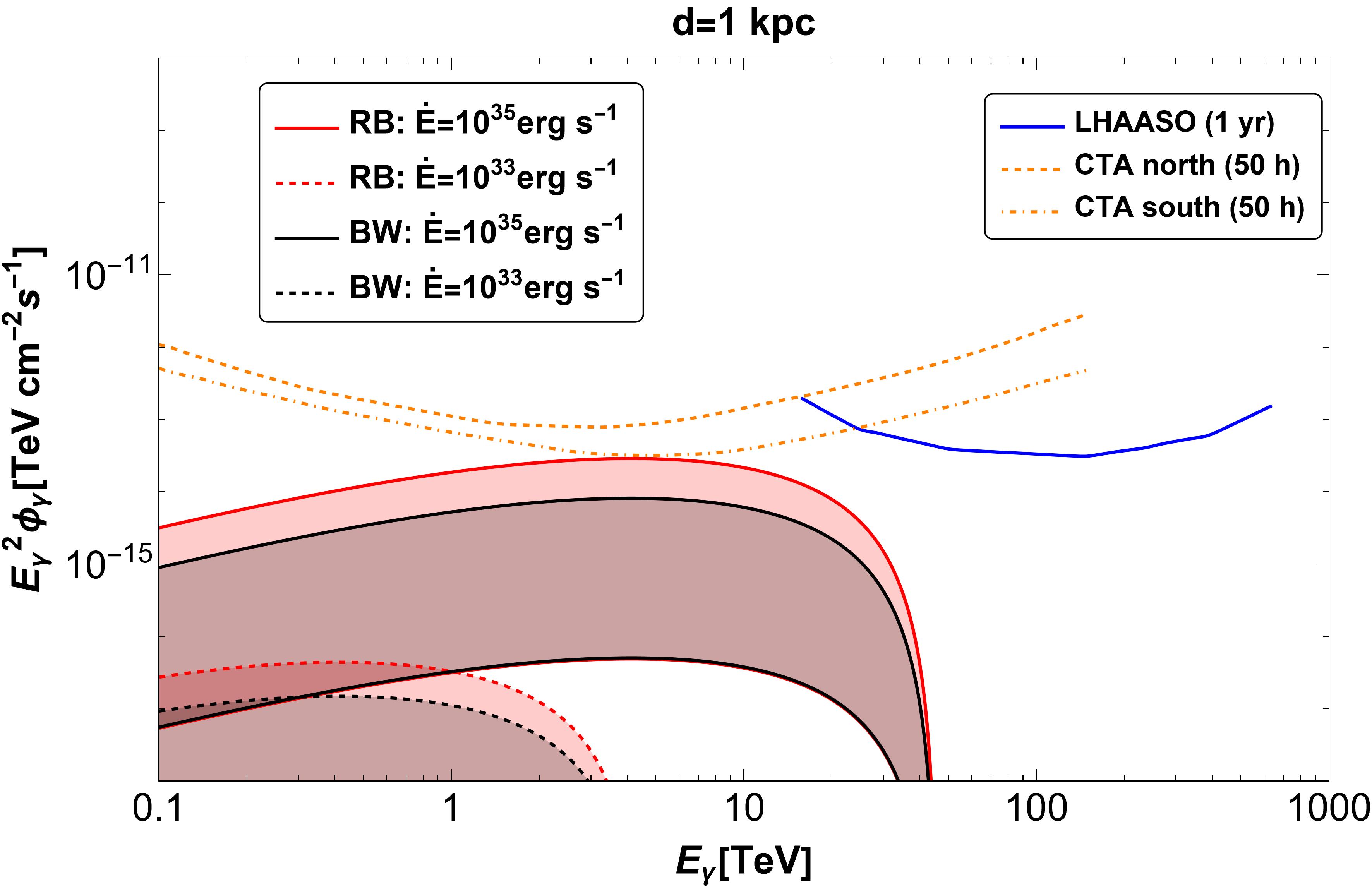}}
\caption{\small\em {\bf Scenario 1: PW$\rightarrow$CW($\gamma$)} Maximum (panel (a)) and minimum (panel (b)) gamma-ray flux from a typical RB (red bands) and BW (black bands). 
All the curves are calculated assuming that protons are injected as a delta function. 
The solid (dashed) lines are obtained for $\dot{E}=10^{35}\,\rm erg\, s^{-1}$ ($\dot{E}=10^{33}\,\rm erg\, s^{-1}$).
The bands include the uncertainty on the magnetic field in the companion region $B_{\rm c}$. The upper (lower) bound is obtained assuming $B_{\rm c}=10^3$ G ($B_{\rm c}=0.1$ G).
The maximum is obtained by calculating the multiplicity according to Eq.~\ref{eq: multiplicity} 
The minimum is obtained assuming a fixed multiplicity of $10^4$.
The dashed and dot-dashed orange lines represent the $50$ h CTA sensitivity in the north and south hemispheres, respectively \citep{Celli:2024cny}.
The blue solid line represents the $1$ year LHAASO sensitivity \citep{Celli:2024cny}.}
\label{fig:Gamma rays}
\end{center}
\end{figure*}

\begin{figure*}
\begin{center}
\subfigure[]{\includegraphics[width=0.45\textwidth]{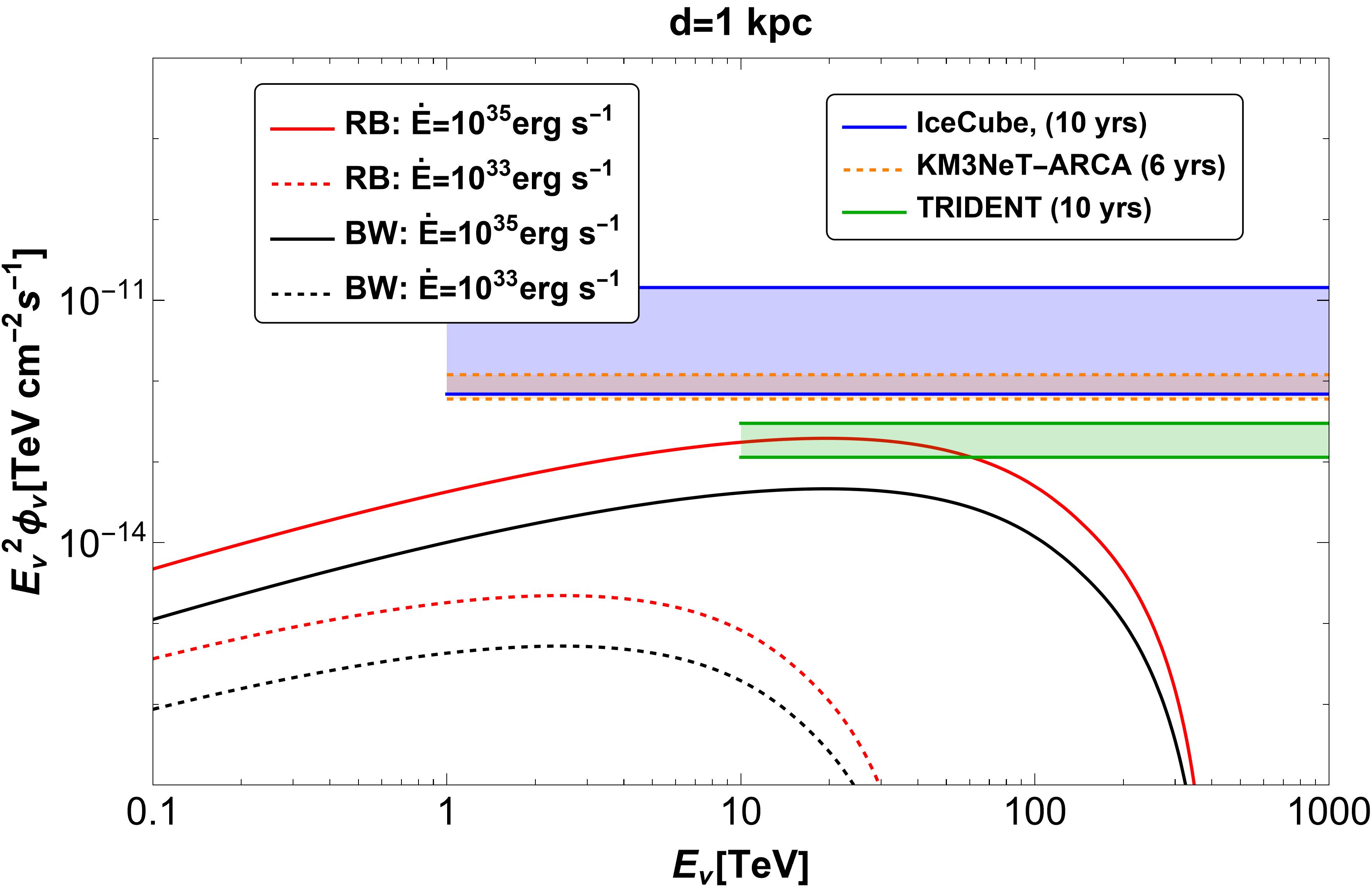}}
\subfigure[]{\includegraphics[width=0.45\textwidth]{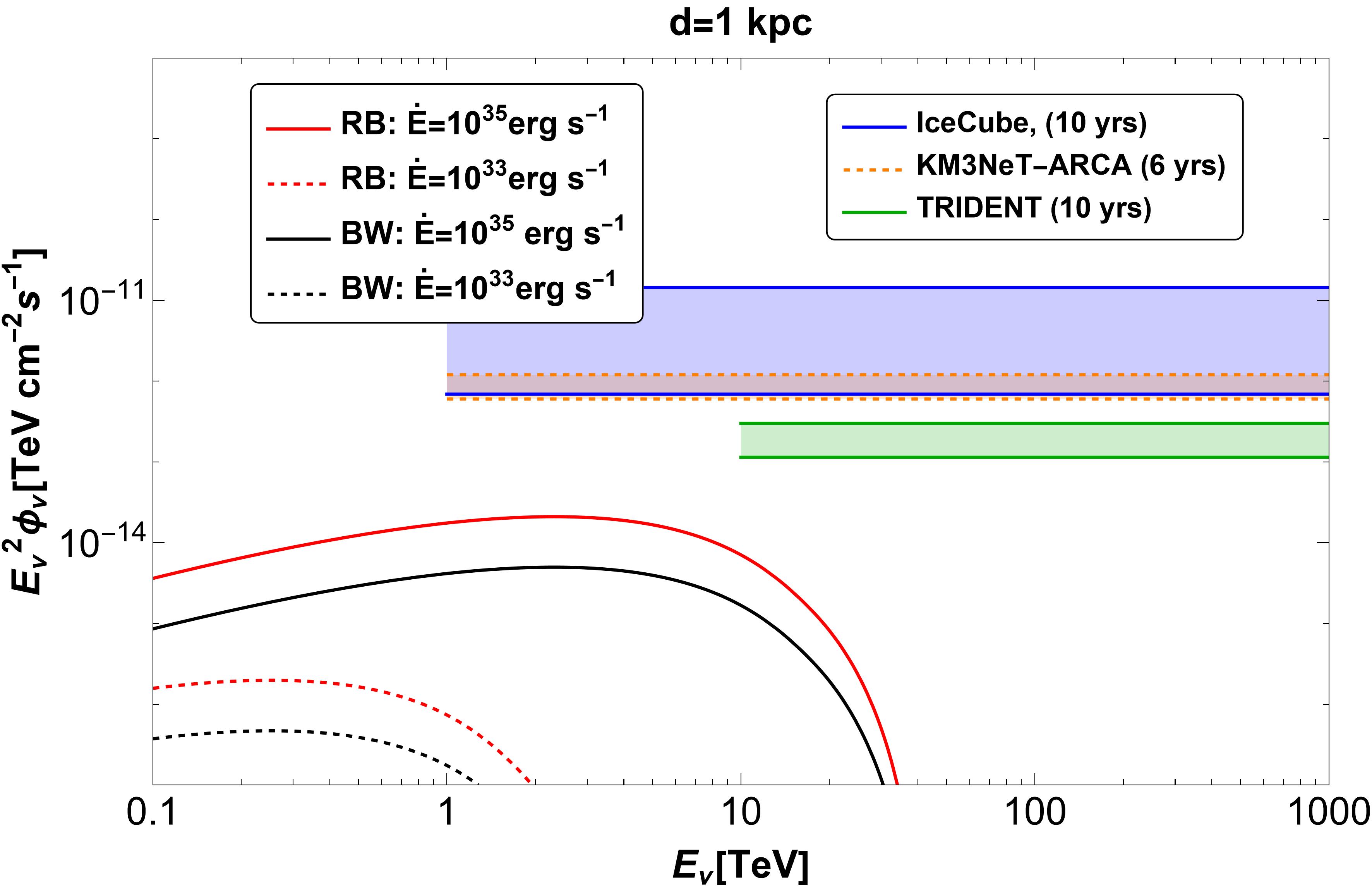}}
\caption{\small\em {\bf Scenario 1: PW$\rightarrow$CS($\nu$)} Same as in Fig.~\ref{fig:Gamma rays} but for the all-flavor neutrino flux.
The $90\%$ confidence-level median sensitivities of the IceCube (blue band) \citep{IceCube:2019cia}, KM3Net-ARCA (orange band) \citep{KM3NeT:2018wnd}, and TRIDENT (green band) \citep{TRIDENT:2022hql} detectors, based on 10, 6, and 10 years of data taking, respectively, are shown assuming a power-law source spectrum with an index of 2.
The exact sensitivity value will depend on the declination of the considered source.}
%The magenta band represents the all-flavors Galactic neutrino flux measured by IceCube on the Galactic plane ($|b|<5^{\circ}$, $|l|<180^{\circ}$) multiplied for the angular region.
\label{fig:Neutrinos}
\end{center}
\end{figure*}

%\subsection{electrons and positrons}

%{\bf estimate using umbrella and estimate in case electrons can cross the shock... maybe we can assume that the fraction of electrons that cross the shock is negligible}

\section{Gamma-ray and neutrino flux}
\label{Sec:Gamma and nu}
We use the result of the above sections to calculate the gamma-ray flux at Earth:
\begin{equation}
\phi_{\gamma}= \frac{c n_{i}}{4 \pi d^2} \left( \frac{S_{i}}{S_{\rm tot}}\right)\int_{E_{\gamma}}^{\infty}dE\frac{d\sigma(E,E_{\gamma})}{dE_{\gamma}}N(E)
\end{equation}
%\begin{equation}
%\phi_{\gamma}= \frac{c n}{4 \pi d^2} \left( \frac{V_{T}}{V_{\rm diff}}\right)\int_{E_{\gamma}}^{\infty}dE'\frac{d\sigma_{\rm pp}(E',E_{\gamma})}{dE_{\gamma}}N(E',t)
%\end{equation}
where $N(E)$ is defined in Eq.~\ref{Eq:solution transport}, $d\sigma(E,E_{\gamma})/dE_{\gamma}$ represents the differential cross section for the production of photons by a nucleon of energy $E$ in a nucleon-nucleon collision, parameterized as in \cite{Kelner:2006tc}, $n_{i}$ is the number density of the wind or the companion and $d$ is the distance of the considered spider source from Earth.
The gamma-ray flux is rescaled by a factor $S_{i}/S_{\rm tot}$.
Since the protons are assumed to be emitted isotropically from the neutron star surface, only a fraction of these protons will intercept the target material.
We define the intercepted surface as $S_{i}$ and the total surface as $S_{\rm tot}=4\pi a^2$. 
In particular, $S_{i}=\pi R_{\rm c}^{2}$ in case of interaction with the CS, while $S_{i}= \pi (4 R_{\rm c})^{2}- \pi R_{\rm c}^{2} =3 \pi R_{\rm c}^{2}$ in case of interaction with the CW. 
%Essentially we are distributing uniformly the protons in a volume $V_{\rm diff}$ and we are making them interact only in the volume occupied by the target $V_{T}$.
%The gamma-ray flux strongly depends on the assumption of $V_{\rm diff}$. We will discuss this in detail later on.
%For scenario 1, the above formula simplifies to:
%
Analogously, we calculate the all-flavor neutrino flux as:
\begin{equation}
\phi_{\nu}= \frac{c n_{i}}{4 \pi d^2} \left( \frac{S_{i}}{S_{\rm tot}}\right)\sum_{l=e,\mu,\tau}\int_{E_{\nu}}^{\infty}dE\frac{d\sigma_{l}(E,E_{\nu})}{dE_{\nu}}N(E)
\end{equation}
where $\frac{d\sigma_{l}(E,E_{\nu})}{dE_{\nu}}$ represents the differential cross section for the production of a neutrino and antineutrino with flavor $l$ by a nucleon of energy $E$ in a nucleon-nucleon collision \citep{Kelner:2006tc}.

%\begin{equation}
%\phi_{\gamma}= \frac{c n_{i}}{4 \pi D^2} \left( \frac{S_{i}}{S_{\rm tot}}\right)\frac{d\sigma_{\rm pp}(E_{\rm C},E_{\gamma})}{dE_{\gamma}}\dot{N}_{\rm GJ}
%\end{equation}

\section{Synthetic population of spiders}
\label{sec: population}
In the final part of this work, we evaluate the contribution of a synthetic population of spiders to the Galactic diffuse emission.
In the following, we perform this calculation only for neutrinos. The gamma-ray diffuse emission measured at very-high energy by LHAASO-KM2A \citep{LHAASO:2023gne} has already been shown to be well explained by hadronic diffuse emission alone, without requiring a significant contribution from unresolved sources \citep{Vecchiotti:2024kkz}. 
%This is primarily due to the masking procedure applied to the data \citep{Vecchiotti:2024kkz}. 

%
In order to estimate the contribution of the entire spider population to the Galactic neutrino flux, we build a synthetic population of these systems. 
For each source, we extract the location in the Galaxy and the spin-down power.
In particular, we assume that the spatial distribution is proportional to the Lorimer distribution \citep{Lorimer:2006qs} expressed in cylindrical coordinates:
\begin{equation}
   n_{\rm s}(\rho,z) = n_{\rm loc}\left(\frac{\rho}{\rho_{\odot}}\right)^{1.9} \exp\left({-b \,\frac{\rho-\rho_{\odot}}{\rho_{\odot}}}\right)\exp\left({-\frac{z}{z_{\rm e}}}\right)
\end{equation}
where $\rho_{\odot}=8.5$ kpc is the distance of the Sun from the Galactic center, $b=5$ and $z_{\rm e}=0.5\,{\rm kpc}$ \citep{Linares:2020fck}.
The distribution is normalized to the local number of spiders $n_{\rm loc}$. This last value is conservatively estimated by assuming that all spiders within a cylinder of $1$ kpc radius have already been detected. 
We adopt the value obtained in \cite{Linares:2020fck}, $n_{\rm loc}\sim 7 \, \rm kpc^{-3}$, as a normalization factor.

The other parameter of our simulation is the spin-down power. 
The histogram of $\dot{E}$ derived using the sample of known spiders (listed in Tab.~\ref{tab:Red Backs} and Tab.~\ref{tab: Black widows}) is reported in Fig.~\ref{fig:histogram}. In general, we expect to be able to detect only the brightest sources, hence the ones that have the highest $\dot{E}$. 
%Hence, we believe that the low spin-down luminosity side of the histogram is not reliable. 
%
Based on Fig.~\ref{fig:histogram}, we consider two possibilities for the logarithmic distribution of the spin-down power. 
The first simple case is a uniform logarithmic distribution for both RBs and BWs. 
In order to do so, we extract uniformly a value in the interval $v=[33,35]$ and calculate the spin-down power as $10^{v}$.
The second possibility is to parameterize the logarithmic distribution of $\dot{E}$ as a Gaussian with mean fixed to $\mu=34.5$ and variance $\sigma=0.5$.
Apart from the location and the spin-down power, all the other parameters are kept constant since they don't strongly affect the final neutrino flux.
We consider 3 cases for scenario 1 (PW$\rightarrow$CS($\nu$)) to account for all possible underlying populations and values of the pair multiplicity $k$:
\begin{enumerate}
\item[a)]  All sources are RB and the pair multiplicity is calculated according to Eq.~\ref{eq: multiplicity}. However, this is a conceptual scenario we consider to maximize the neutrino flux, but it doesn't hold up in reality;
\item[b)]  Half of the sources are RBs, and the other half are BWs. The pair multiplicity is calculated according to Eq.~\ref{eq: multiplicity};
\item[c)]  Half of the sources are RBs, and the other half are BWs. The pair multiplicity is fixed to $k=10^{4}$;
\end{enumerate}
For each case, we simulate 100 populations where each population has an average number of $\sim 3600$ spiders, obtained by integrating the spatial distribution in the region $|l|<180^{\circ}$ and $|b|<5^{\circ}$, where IceCube reported the measurement of the Galactic diffuse neutrino emission \citep{IceCubeScience}.

\begin{figure*}
\begin{center}
\subfigure[]{\includegraphics[width=0.45\textwidth]{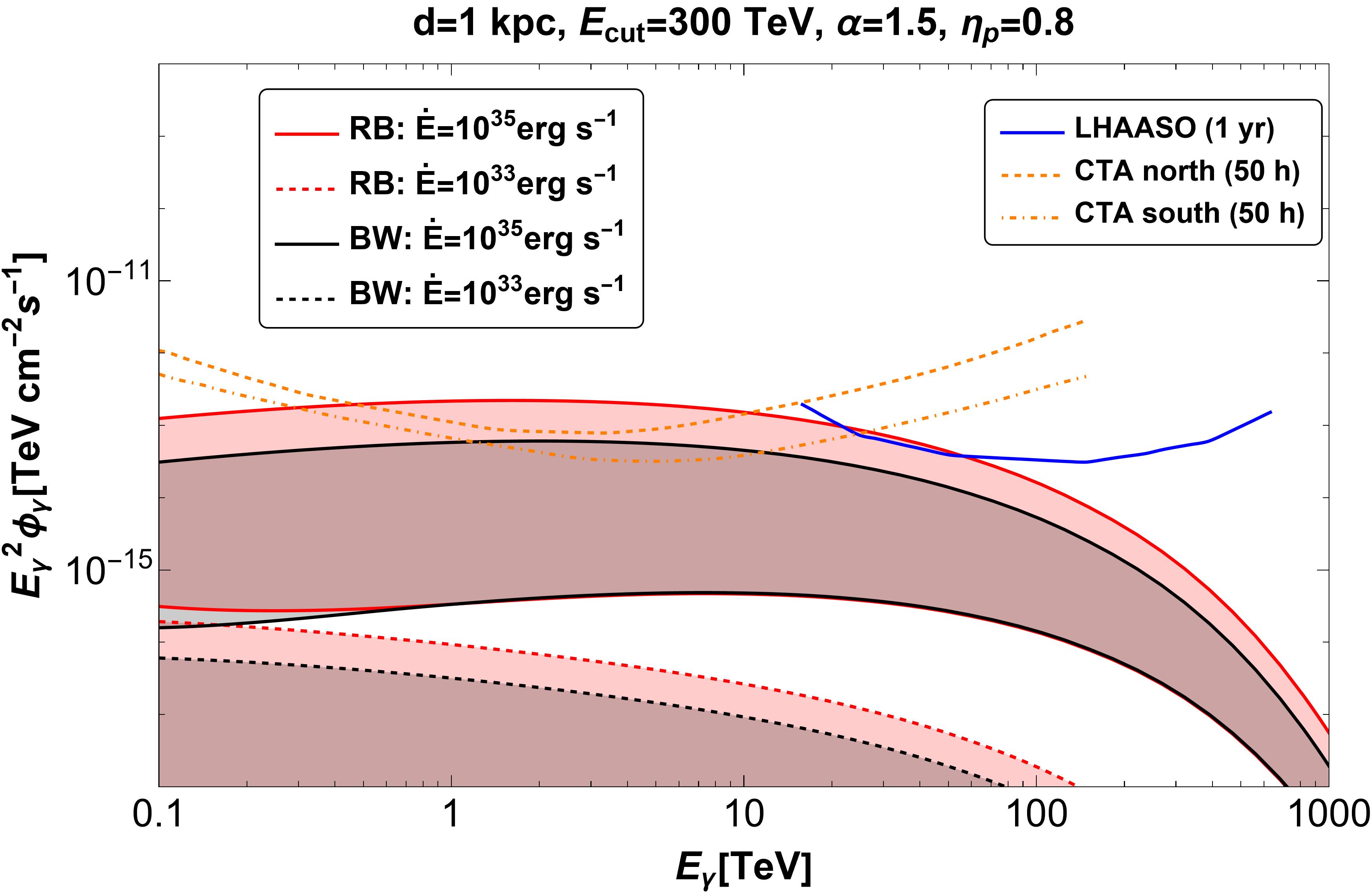}}
\subfigure[]{\includegraphics[width=0.45\textwidth]{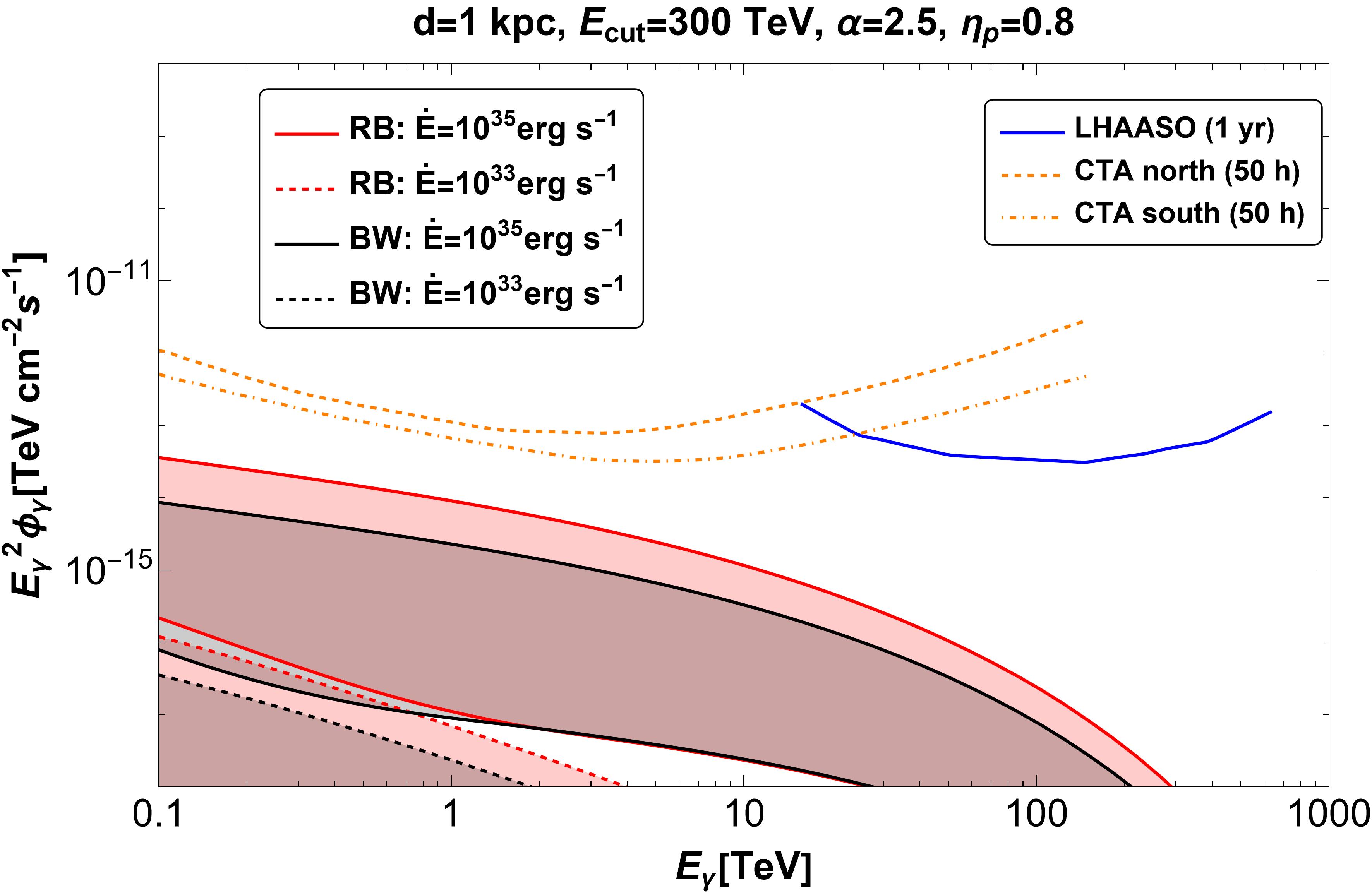}}
\caption{\small\em  {\bf Scenario 2: IBS$\rightarrow$CW($\gamma$)} Maximum for gamma-ray flux from a typical RB (red lines) and BW (black lines). 
All the curves are calculated assuming $\eta_{\rm p}=0.8$ and that protons are injected as a power-law with an exponential cut-off, where the spectral index is fixed to $\alpha=1.5$ (panel (a)) and $\alpha=2.5$ (panel (b)) and the cut-off energy is fixed to $300$ TeV, calculated according to Eq.~\ref{eq: Ecut}.
The bands include the uncertainty on the magnetic field in the companion region $B_{\rm c}$. The upper (lower) bound is obtained assuming $B_{\rm c}=10^3$ G ($B_{\rm c}=0.1$ G).
}
\label{fig:Gamma ray PL}
\end{center}
\end{figure*}

\begin{figure*}
\begin{center}
\subfigure[]{\includegraphics[width=0.45\textwidth]{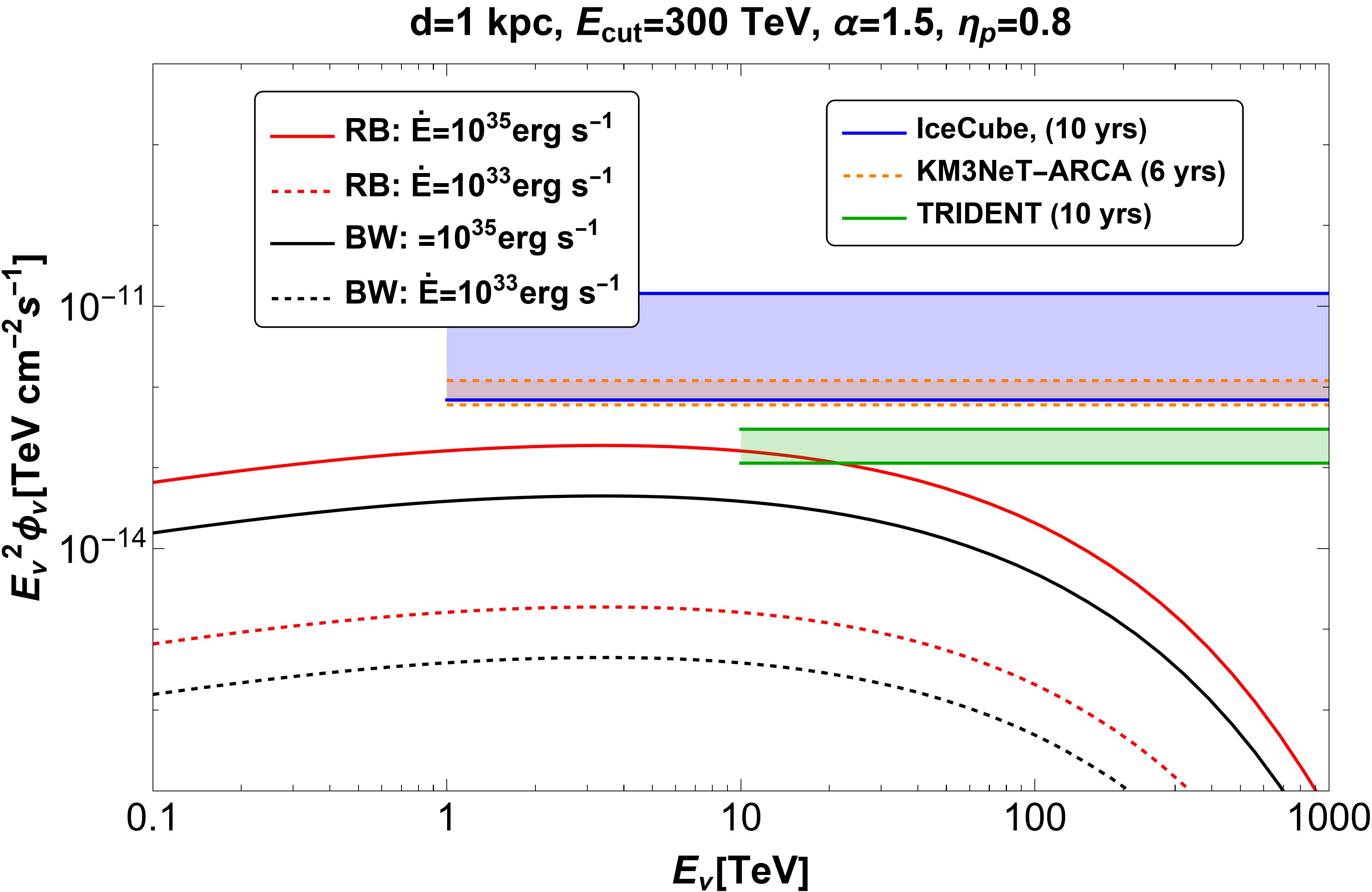}}
\subfigure[]{\includegraphics[width=0.45\textwidth]{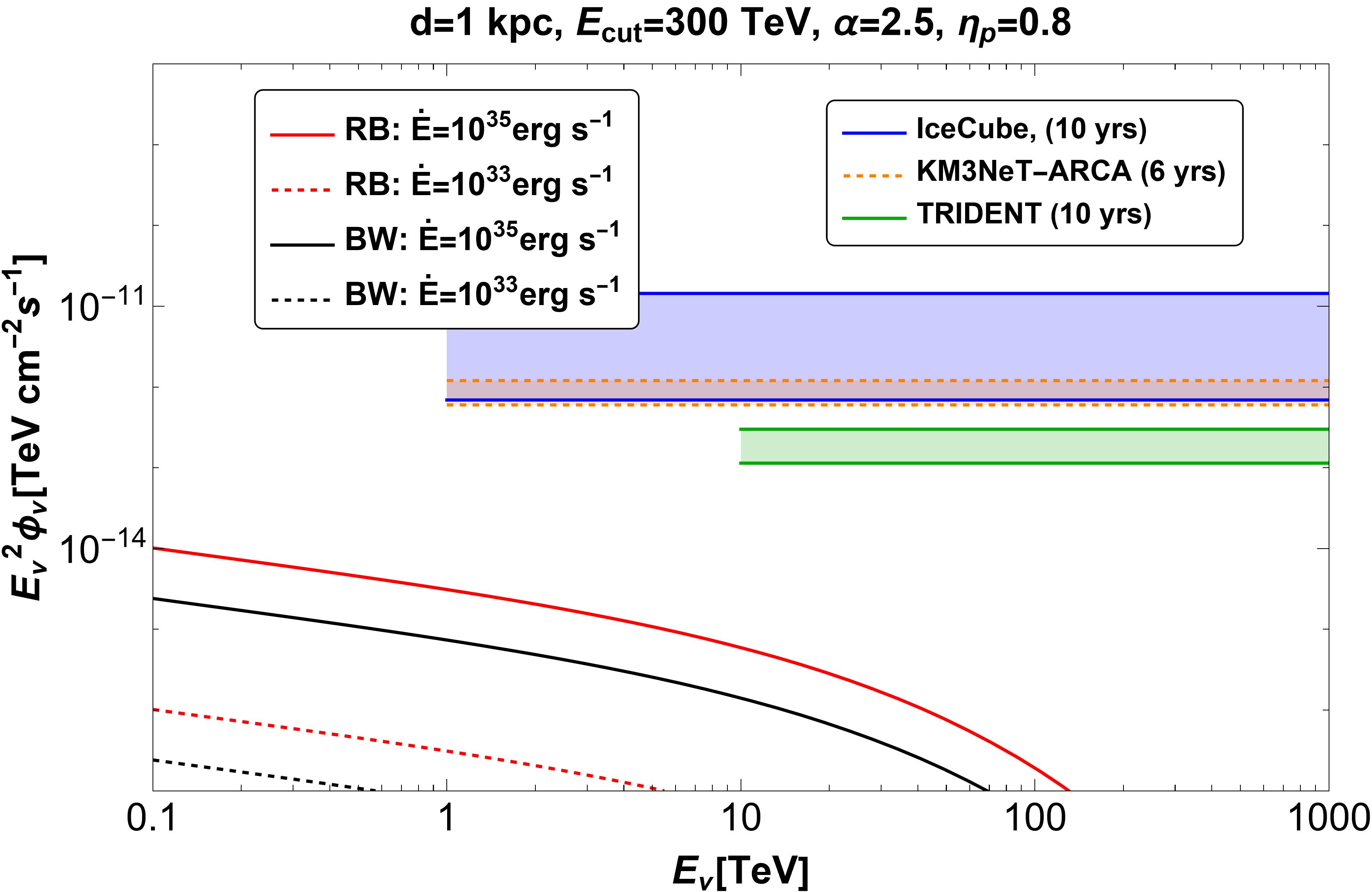}}
\caption{\small\em {\bf Scenario 2: IBS$\rightarrow$CS($\nu$)} Same as in Fig.~\ref{fig:Gamma ray PL} but for the all-flavor neutrino flux.
The $90\%$ confidence-level median sensitivities of the IceCube (blue band) \citep{IceCube:2019cia}, KM3Net-ARCA (orange band) \citep{KM3NeT:2018wnd}, and TRIDENT (green band) \citep{TRIDENT:2022hql} detectors, based on 10, 6, and 10 years of data taking, respectively, are shown assuming a power-law source spectrum with an index of 2.
}
%The blue line represents the IceCube sensitivity for Cascades \vitt{southern and northern Hemisphere} \citep{}.
%
%The dashed orange lines represent the $10$ yr KM3NeT-ARCA sensitivity curve in the Southern Hemisphere \citep{KM3NeT:2018wnd}.
%The magenta band represents the all-flavors Galactic neutrino flux measured by IceCube on the Galactic plane ($|b|<5^{\circ}$, $|l|<180^{\circ}$) multiplied for the angular region.
\label{fig:Neutrinos PL}
\end{center}
\end{figure*}

\section{Results}
\label{sec:Result}

\subsection{Generic single spiders}
\label{sec: Single spiders}
Using typical values for RBs and BWs, as listed in Tab.~\ref{tab: standard parameters}, we calculate the gamma-ray and the neutrino emission produced by the interaction of protons with the CW and the CS, respectively.
We predict the fluxes for the two proton acceleration scenarios considered in this work (PW and IBS).

%scenario 1
In Figures \ref{fig:Gamma rays} and \ref{fig:Neutrinos}, we show the gamma-ray and all-flavor neutrino flux in the case of scenario 1 (PW),
%flux from a typical RB (red bands) and BW (black bands) 
for two different assumptions on the spin-down power: $\dot{E}=10^{33}\, \rm erg^{-1}\,s^{-1}$ (dashed lines) and $\dot{E}=10^{35}\, \rm erg^{-1}\,s^{-1}$ (solid lines).
All the curves are calculated assuming that protons are injected as a delta function, with the predicted maximum achievable flux displayed in panel (a) and the minimum in panel (b). 
In particular, the maximum for the gamma-ray and neutrino fluxes is obtained by calculating the pair multiplicity $k$ according to Eq.~\ref{eq: multiplicity}. 
The minimum flux is instead obtained assuming a fixed pair multiplicity of $10^4$.

%scenario 2
We display our prediction for the gamma-ray and all-flavor neutrino in the case of scenario 2 (IBS) in Figures \ref{fig:Gamma ray PL} and \ref{fig:Neutrinos PL}, respectively.
%The results are again shown for two different assumptions on the spin-down power: $\dot{E}=10^{33}\, \rm erg^{-1}\,s^{-1}$ (dashed lines) and $\dot{E}=10^{35}\, \rm erg^{-1}\,s^{-1}$ (solid lines).
%
All the curves are calculated assuming that protons are injected as a power-law with an exponential cut-off.
The spectral index is fixed to $\alpha=1.5$ in panel (a) and $\alpha=2.5$ in panel (b), while the cut-off energy is fixed at $300$ TeV, as calculated from Eq.~\ref{eq: Ecut} for the highest magnetization $\sigma$ achievable in our model.
We have also fixed the fraction of spin-down power transferred to protons to the maximum $\eta_{\rm p}=0.8$. For scenario 2 (IBS), we chose to display only our prediction for the maximum flux, since changing $\eta_{\rm p}$ in this case simply rescales the normalization without affecting the maximum achievable energy, which is instead determined by the magnetization in the upstream region of the IBS.

% gamma-rays
Looking at Figures \ref{fig:Gamma rays} and \ref{fig:Gamma ray PL}, we notice that, for specific parameter combinations, a spider system could produce a gamma-ray flux that is above the sensitivity threshold of CTA (orange lines) and LHAASO (blue line) \citep{Celli:2024cny}.
In particular, in scenario 1 (PW) for a RB-like system, if the spider has a spin-down power of $\dot{E}=10^{35}\,\rm erg\, s^{-1}$ and it is located at a distance $d=1\,\rm kpc$ from the Sun, the source could be detected by both LHAASO and CTA, provided the magnetic field in the companion region is sufficiently high and the multiplicity is sufficiently low (see panel (a) of Fig.~\ref{fig:Gamma rays}).
% However, if the multiplicity is fixed to $10^4$, the source could be detected only by CTA (see panel (b) of Fig.~\ref{fig:Gamma rays}), assuming the same spin-down power and high magnetic field.
However, if the multiplicity is fixed to $10^4$, the source would barely be detectable by CTA (see panel (b) of Fig.~\ref{fig:Gamma rays}), assuming the same spin-down power and high magnetic field.
Spiders at distances shorter than 1~kpc would have higher gamma-ray fluxes \citep[e.g., the RB candidate 3FGL J0737.2-3233 at $d=0.66\pm0.02\,\rm kpc$,][]{Turchetta24}.

In scenario 2 (IBS), the requirements for detecting a RB-like system in terms of spin-down power and magnetic field are similar to those in scenario 1 (PW).
% Interestingly, for the most optimistic case with $\eta_{\rm p}=0.8$ and $\alpha=1.5$, the signal exceeds the detection thresholds of both CTA and LHAASO KM2A.
%However, since the cut-off energy of the injected spectrum is fixed at 300 TeV, most of the gamma rays are produced within the CTA energy range rather than in that of LHAASO KM2A.
In the most optimistic case with $\eta_{\rm p}=0.8$ and $\alpha=1.5$, the signal exceeds only the detection thresholds of CTA and the low-energy end of the LHAASO's threshold.
This is due to the cut-off energy of the injected spectrum, which is fixed at $300$ TeV; as a result, most of the gamma rays are produced within the CTA energy range, rather than that of LHAASO KM2A.
It is also important to note that for lower values of the magnetization $\sigma$, the resulting cut-off energy would be lower, making detection with the LHAASO-KM2A detector unlikely.
The injected spectral index is also a crucial parameter.  In fact, if the injected proton spectrum is too soft, for example, such as with $\alpha=2.5$ (as shown in panel (b) of Fig.~\ref{fig:Gamma ray PL}), the IBS-CW signal would always remain below the detection threshold.
%Therefore, the detectability of spider sources in this scenario depends on the acceleration mechanism at the IBS. If diffusive shock acceleration takes place, the accelerated proton spectrum could be too soft to produce a detectable signal, even under the most favorable conditions.
%On the other hand, if particles are accelerated via magnetic reconnection, the injected proton spectrum might be harder than $\alpha \sim 2$, which could result in a detectable signal for CTA.

%
In general, the main caveat of above considerations is that, for low values of the pair multiplicity $k$ (in scenario 1: PW) and low values of the fraction of spin-down power transferred to protons $\eta_{\rm p}$ (in scenario 2: IBS), spider sources are unlikely to produce a detectable signal via the two suggested scenarios for current and upcoming gamma-ray detectors.
%{\color{magenta} instead of realistic put "low". k=100-133 from H and M is state of the art}
%Especially looking at known systems could be used to constrain the magnetic field of the companion

% neutrinos
In the neutrino case (see Fig.~\ref{fig:Neutrinos}), a nearby spider system could produce a flux above 
%the $90\%$ confidence level median sensitivity of KM3Net-ARCA (orange band) \citep{KM3NeT:2018wnd} and 
the upcoming TRIDENT (green band) \citep{TRIDENT:2022hql} detectors in Scenario 1 (PW-CS) for the most optimistic case (see panel (a) of Fig.~\ref{fig:Neutrinos}).
In the case of scenario 2 (IBS-CS), shown in Fig.~\ref{fig:Neutrinos PL}, an individual system could produce a detectable signal only above the low-energy end of the TRIDENT threshold, and only for $\alpha=1.5$.
%In the case of scenario 2 (IBS-CS), shown in Fig.~\ref{fig:Neutrinos PL}, the signal barely reaches the sensitivity threshold of current and future neutrino detectors for the case with $\alpha=1.5$, while no individual system produces a detectable signal for $\alpha=2.5$.
% This is mainly due to the lower cut-off in the proton energy, $E_{\rm cut}= 300$ TeV, compared to scenario 1, which results in a cut-off in the neutrino energy at $\sim  15 $ TeV.

% 
Although it seems challenging to detect the neutrino signal from single spiders, the cumulative neutrino signal produced by all the spider systems in our Galaxy could provide a non-negligible contribution to the measured IceCube signal.
In the following, we quantify this contribution.

%In general, these sources produce a gamma ray signal that exhibits a peck in the energy range $1-10^3$ TeV, especially in the region where LHAASO-KM2A is more sensitive.

%We select all the known CBMP in the LHAASO observation windows and we calculate the expected gamma-ray signal produced by accelerated protons interacting with the companion star.
%We show that some systems produce a signal very close to the LHAASO-KM2A detection sensitivity, implying that if protons are extracted from the pulsar surface we should be able to see these systems with LHAASO-KM2A.
%The IC from pairs cannot produce a peak at such high energy, due to strong synchrotron losses.
%As a consequence, the observation of these sources with LHAASO would prove that pulsars can accelerate protons up to very high energy.

\subsection{Sample of known spiders}

For the most optimistic scenario 1 (PW) and assuming $k$ from Eq.~\ref{eq: multiplicity} to maximize the flux, we calculate the gamma-ray and neutrino flux from a sample of known spiders with measured distance and spin-down power.
The sample considered includes $19$ RBs and $32$ BWs, which constitute the subset of systems for which we have information on the dispersion measure distance $d_{\rm DM}$ and spin-down power $\dot{E}$, as reported in Tables \ref{tab:Red Backs} and \ref{tab: Black widows} \citep{Nedreaas2024}.

\begin{figure*}
\begin{center}
\subfigure[]{\includegraphics[width=0.45\textwidth]{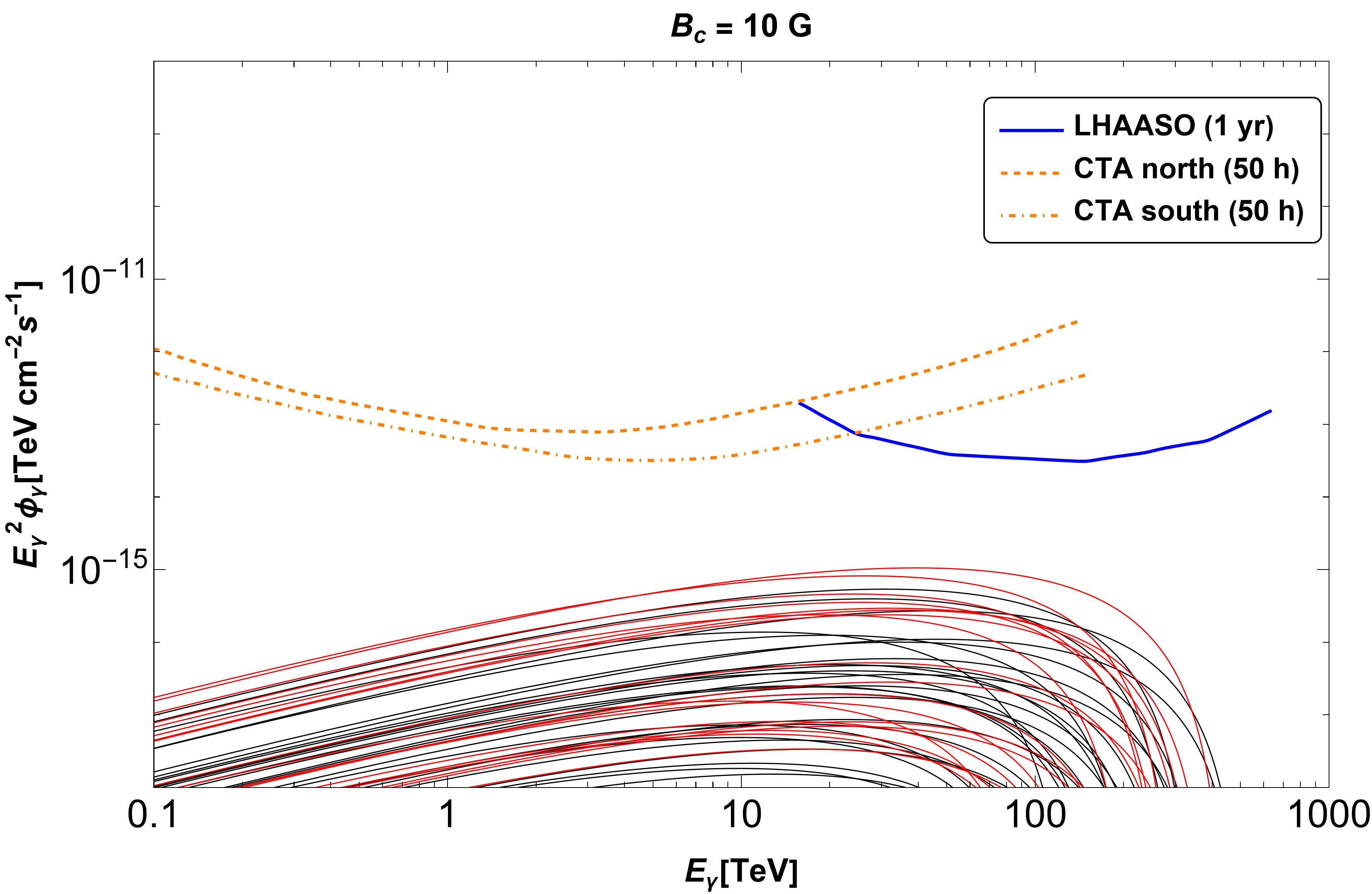}}
\subfigure[]{\includegraphics[width=0.45\textwidth]{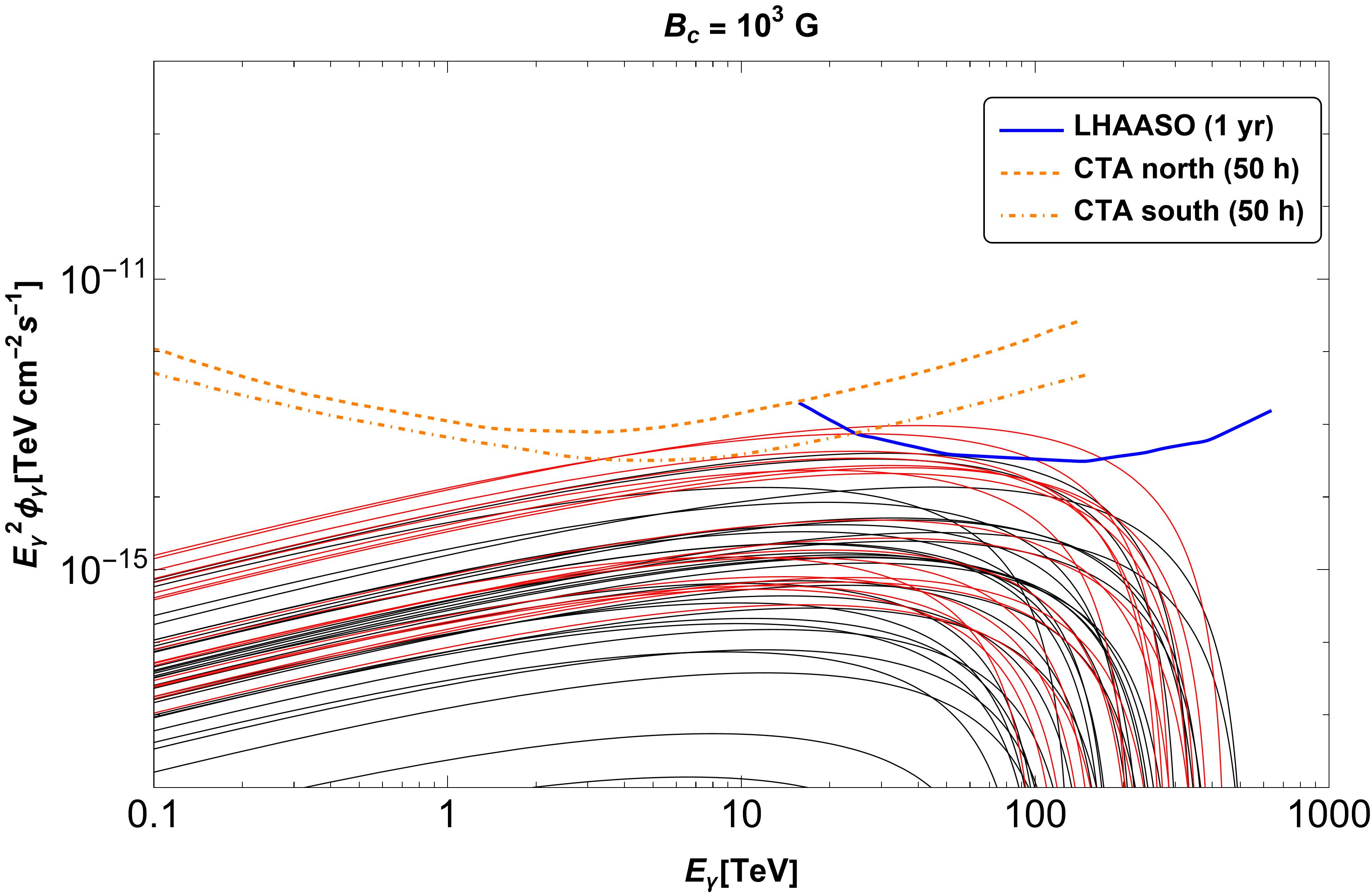}}
\caption{\small\em  Panel (a): Gamma-ray flux from known RBs (red thin lines) and BWs (black thin lines), listed in Tab.~\ref{tab:Red Backs} and Tab.~\ref{tab: Black widows}, respectively. The flux is calculated assuming a companion magnetic field fixed at $B_{\rm c}=10$ G. We also display the LHAASO $1$ yr sensitivity threshold (blue line) and the $50$ h CTA sensitivity threshold (orange lines).
Panel (b): Same as panel (a) but with companion magnetic field fixed at $B_{\rm c}=10^3$ G.
All the curves are calculated assuming scenario 1 with $k$ calculated from Eq.~\ref{eq: multiplicity}.
%, overlapped to our prediction for the maximum gamma-ray flux (panel (a) Fig.~\ref{fig:Gamma rays}).
}
\label{fig:known spiders gamma}
\end{center}
\end{figure*}

\begin{figure}
%\begin{center}
\includegraphics[width=0.45\textwidth]{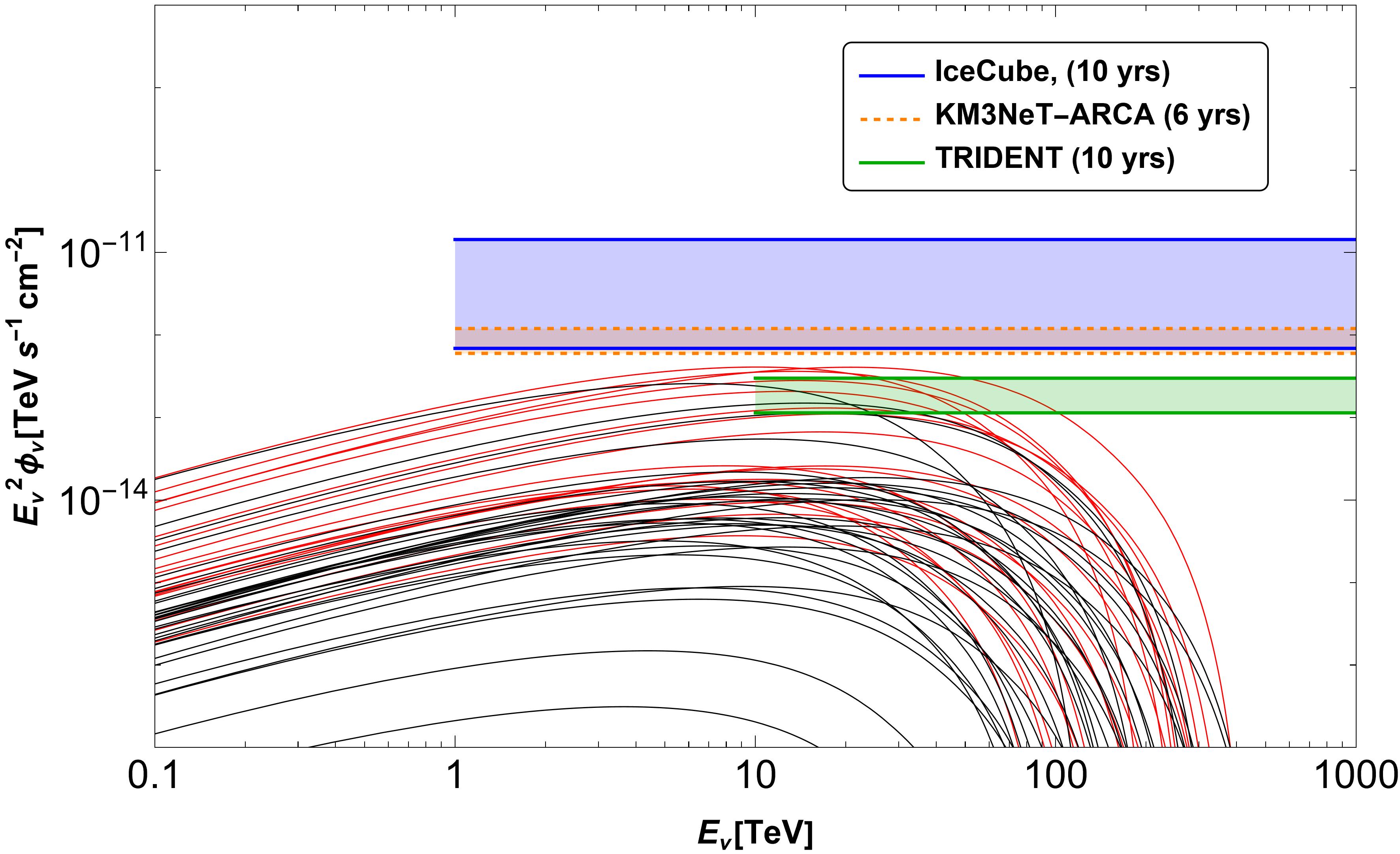}
\caption{\small\em 
All-flavour neutrino flux from known RBs (red thin lines) and BWs (black thin lines).
We also display the $10$ yrs IceCube (blue line) and KM3Net (dashed orange line) sensitivity threshold.
All the curves are calculated assuming scenario 1 with $k$ calculated from Eq.~\ref{eq: multiplicity}.}
\label{fig:known spiders nu}
%\end{center}
\end{figure}

The results of our calculation are shown in Figures \ref{fig:known spiders gamma} and \ref{fig:known spiders nu} for gamma-rays and neutrinos, respectively.
In particular, panel (a) of  Fig.~\ref{fig:known spiders gamma}, displays the results assuming a companion magnetic field of 10 G, while panel (b), shows the result for a companion magnetic field of $10^{3}$ G.
As noted in Sec.~\ref{sec: timescales}, this parameter plays a crucial role in gamma-ray predictions. 
A high value of $B_{\rm c}$ could make some spider systems detectable by LHAASO and the upcoming CTA.
This is especially true for two RB systems, PSR J1910-5320 and PSR J1723-2837, which appear to produce a signal above the LHAASO sensitivity threshold (see panel (b) of Fig.~\ref{fig:known spiders gamma}).
These are two nearby sources ($d_{\rm DM}\le1$ kpc) with high spin-down powers of $\sim 10^{35}\, erg\,s^{-1}$.
A counterpart for PSR J1910-5320 has been identified in the 4FGL catalog \citep{Fermi-LAT:2019yla}, precisely the Fermi-LAT sources 4FGL J1910.7-5320 \citep{Au:2022ole,Dodge:2024nvl}.
%while the system {\bf PSR} J1723-2837 has not been detected as a gamma-ray source by Fermi-LAT \citep{Crawford:2013hoa}.
PSR J1723-2837,  on the other hand, shows some hint of gamma-rays, as reported by \cite{Hui:2013tka} and \cite{vanderMerwe2020ApJ}.
%However, the LHAASO collaboration, in their first catalog based on $508$ days of data collected by the Water Cherenkov Detector Array (WCDA) and the Kilometer Squared Array (KM2A), has not yet reported any significant detections from the locations of these two systems \citep{LHAASO:2023rpg}.
%This measurement does not exclude the possibility that, with more data, it may still be possible to detect these spiders.
%In general, for the above-mentioned systems, we can set an upper limit on the magnetic field in the companion region: it must be $< 10^{3}$ G.
However, due to their large zenith angles ($z > 50^{\circ}$), these sources fall outside LHAASO’s observational window \citep{LHAASO:2023rpg}.

Regarding neutrinos (see Fig.~\ref{fig:known spiders nu}),  none of the currently known spiders are expected to produce a detectable signal for IceCube \citep{IceCube:2019cia} or the KM3NeT-ARCA \citep{KM3NeT:2018wnd}.
However, the future TRIDENT detector \citep{TRIDENT:2022hql}, with its improved sensitivity, could potentially detect neutrino emission from a group of spider sources, particularly RBs.
In general, it is worthwhile to investigate the contribution of the cumulative neutrino flux from a synthetic population of spider sources to the currently measured diffuse neutrino emission \citep{IceCubeScience}.
% Although it is challenging to detect neutrinos from single sources, in the next section we investigate the cumulative flux from a synthetic population of spider sources.

%In Sec.~\ref{sec: Single spiders}, we demonstrated that for specific parameter combinations such as the distance from the Sun $d$, spin-down power $\dot{E}$ and magnetic field in the companion region $B_{\rm c}$, some system should already have been detected by LHAASO.
%Since no spider has yet been detected by LHAASO, we can use our prediction for gamma-rays from known systems, listed in Tab.~\ref{tab:Red Backs} and Tab.~\ref{tab: Black widows}, to set an upper limit on the magnetic field in the companion region.

\subsection{Cumulative neutrino flux}
\begin{figure}
\includegraphics[width=0.45\textwidth]{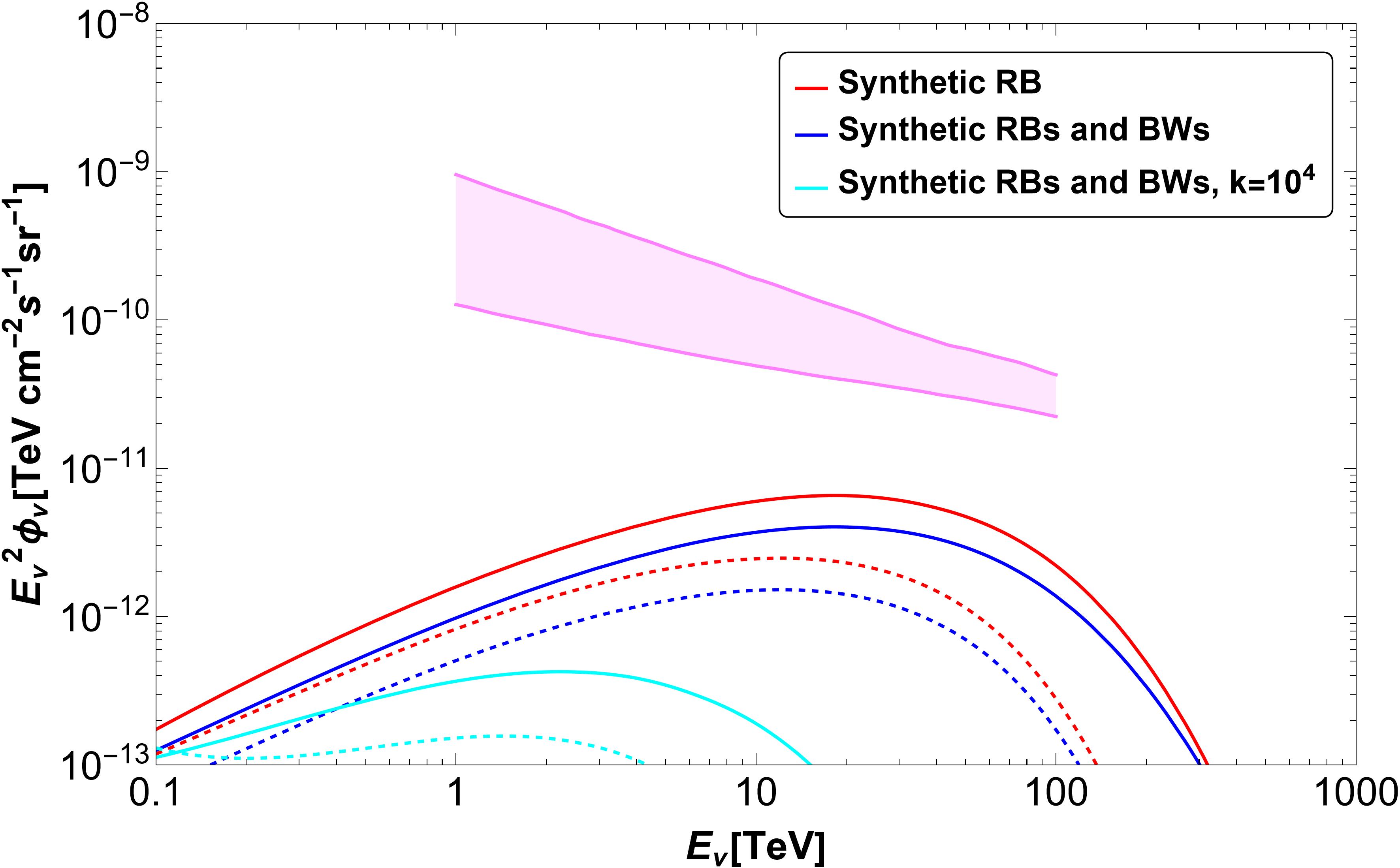}
\caption{\small\em  
Differential energy spectra of the all-flavor neutrino flux from a population of spiders, all RBs (red lines), half RBs and half BWs (blue lines), half RBs and half BWs with pair multiplicity $k=10^4$ (cyan lines) in the sky region $|l|\le 180^{\circ}$ and $|b|\le 5^{\circ}$. 
Solid lines are obtained by extracting $\log_{10}\dot{E}$ from a Gaussian distribution with mean $\mu=34.5$ and variance $\sigma=0.5$.
Dashed lines are obtained by extracting $\log_{10}\dot{E}$ from a uniform distribution in the interval $[33,35]$.
The magenta region corresponds to the superposition of the three IceCube best-fits for the galactic component with their $1\sigma$ uncertainty \citep{IceCubeScience}.
}
\label{fig: Nu from Population}
\end{figure}

In Fig.~\ref{fig: Nu from Population}, we show the total neutrino flux produced by our synthetic population of spiders for the 3 cases listed in Sec.~\ref{sec: population} in the sky region: $|l|\le 180^{\circ}$ and $|b|\le 5^{\circ}$. The total flux is calculated as the median value over 100 realizations.
We present the median value instead of the average to reduce the impact of large fluctuations, which can arise in some realizations due to the occasional presence of a powerful source located near the Sun.
The dashed lines are calculated assuming that $\log_{10}\dot{E}$ is distributed uniformly, while the solid lines are obtained assuming a Gaussian distribution. The former possibility produces slightly lower fluxes with respect to the latter one.

Our predictions for Case a) are displayed with red lines. This case, though unrealistic, produces the largest neutrino flux. Indeed, the total neutrino flux from the spider population calculated assuming a $\log_{10}\dot{E}$ Gaussian (uniform) distribution can contribute up to $5-16\%$ ($2-6\%$) to the IceCube diffuse emission flux at $20$ TeV \cite{IceCubeScience}.
In Case b) (see blue lines), obtained by considering a combined population of RBs and BWs, the percentage decreases to $3-10\%$ ($1-3\%$).
%However, it assumes a large value of $\eta_{p}\sim 0.8$. 

%
Lastly, in Case c) (see cyan lines), with $k=10^4$ and consequently $\eta_{p}\sim 0.08$, it produces a neutrino flux that is a few orders of magnitude below the level of the Galactic diffuse emission measured by IceCube \citep{IceCubeScience}.

In general, including BWs in the spider population has the effect of slightly decreasing both the maximum achievable energy and the total flux.
Increasing the multiplicity $k$, as already noted in Sec.~\ref{sec: multiplicity}, has the same effect in a more drastic way.

In conclusion, we find that the total population of spiders gives a negligible contribution to the total galactic neutrino flux measured by IceCube.

\section{Conclusions}
\label{Sec:Conclusion}

In this paper, we have calculated the very high-energy (0.1–$10^3$ TeV) gamma-ray and neutrino emission in spider systems, produced by protons accelerated at the IBS and/or within the PW as they interact with the CW and the CS.
We have shown that the interaction between the IBS and CW can produce detectable gamma-ray emission at $\sim$1 TeV for CTA, but only if the injected proton spectrum at the shock is hard ($\alpha = 1.5$) and the companion’s magnetic field is sufficiently strong ($\sim 10^3$ G).
We also find that the interaction between the IBS and CS is unlikely to generate detectable neutrino emission.
On the other hand, the interaction between the PW and CW may produce a detectable gamma-ray signal at a few TeV (CTA) and possibly at $\sim$100 TeV (LHAASO), provided the pair multiplicity is low ($k \sim 100–300$). 
Regarding neutrinos, the interaction between the PW and CS could lead to detectable neutrino emission for the future TRIDENT detector \citep{TRIDENT:2022hql}, again assuming a low pair multiplicity.
Finally, we have estimated the contribution of a synthetic population of spiders to the diffuse neutrino flux recently measured by IceCube \citep{IceCubeScience}, showing that this contribution is negligible.

\begin{acknowledgments}
The authors are grateful to Alice K. Harding, Elena Amato, Benedikt Schroer, and Foteini Oikonomou for their helpful discussions and comments.
The work of VV and ML is supported by the European Research Council (ERC) under the ERC-2020-COG ERC Consolidator Grant (Grant agreement No.101002352).
\end{acknowledgments}

\bibliography{bibliography.bib}
\bibliographystyle{aasjournalv7}

\begin{table*}
\centering
\tablenum{2}
\begin{tabular}{|c|c|c|c|c|c|}
\hline
\textbf{Name} & \textbf{dist (kpc)} & \textbf{$d_{DM}$ (kpc)} & \textbf{$\dot{E}(\times 10^{34}\, erg \, s^{-1})$} & \textbf{$P_{b}$ (h)} & \textbf{$M_{c} (M_{\odot})$} \\ \hline
PSR J0212+5320      & 1.1           & 1.27          & -             & 20.9          & 0.3           \\ \hline
PSR J1023+0038      & 1.37          & 0.6           & 4.3           & 4.8           & 0.2           \\ \hline
PSR J1036-4353   & -             & 0.4           & -             & 6.2           & 0.23          \\ \hline
PSR J1048+2339    & -             & 1.7           & 1.2           & 6.0           & 0.3           \\ \hline
PSR J1227-4853      & 2.5           & 1.4           & 9.0           & 6.9           & 0.14          \\ \hline
PSR J1302-3258  & -             & 0.96          & 0.5           & 15.4          & 0.15          \\ \hline
PSR J1306-40        & 4.7           & 1.2           & -             & 26.3          & -             \\ \hline
PSR J1431-4715   & -             & 1.5           & 6.8           & 10.8          & 0.12          \\ \hline
PSR J1526-2744 & -             & 1.3           & 0.91          & 4.9           & 0.083         \\ \hline
PSR J1622-0315      & -             & 1.1           & 0.77          & 3.9           & 0.1           \\ \hline
PSR J1628-3205    & 1.2           & 1.2           & 1.35          & 5.0           & 0.16          \\ \hline
PSR J1723-2837      & 0.75          & 0.75          & 4.7           & 14.8          & 0.24          \\ \hline
PSR J1803-6707   & -             & 1.4           & 7.4           & 9.1           & 0.26          \\ \hline
PSR J1816+4510    & 4.5           & 2.4           & 5.2           & 8.7           & 0.16          \\ \hline
PSR J1908+2105   & 2.6           & 2.6           & 3.2           & 3.5           & 0.055         \\ \hline
PSR J1910-5320   & 4.4           & 1.0           & 11.5          & 8.4           & 0.28          \\ \hline
PSR J1957+2516  & -             & 3.1           & 1.7           & 5.7           & 0.1           \\ \hline
PSR J2039-5618   & 1.7           & 1.7           & 2.5           & 5.4           & -             \\ \hline
PSR J2055+1545      & -             & 3.7           & 3.83          & 4.82          & 0.24          \\ \hline
PSR J2129-0429    & 0.9           & 0.9           & 3.9           & 15.2          & 0.37          \\ \hline
PSR J2215+5135      & 3.0           & 3.0           & 7.4           & 4.1           & 0.22          \\ \hline
PSR J2339-0533      & 1.1           & 0.45          & 2.3           & 4.6           & 0.26          \\ \hline
PSR J0407.7-5702 & 7.0           & -             & -             & -             & -             \\ \hline
PSR J0427.9-6704 & 2.4           & -             & -             & 8.8           & -             \\ \hline
PSR J0523-2529   & 1.1           & -             & -             & 16.5          & 0.8           \\ \hline
PSR J0744-2523 & 1.5           & -             & -             & 2.8           & -             \\ \hline
PSR J0838.8-2829 & 1.0         & -             & -             & 5.1           & -             \\ \hline
PSR J0935.3+0901 & -             & -             & -             & 2.44          & -             \\ \hline
PSR J0940.3-7610 & 2.0           & -             & -             & 6.5           & -             \\ \hline
PSR J0954.8-3948 & 1.7           & -             & -             & 9.3           & -             \\ \hline
PSR J1544-1128     & 3.8           & -             & -             & 5.8           & -             \\ \hline
PSR J1646.5-4406 & -             & -             & -             & 5.27          & -             \\ \hline
PSR J1702.7-5655 & -             & -             & -             & 5.85          & -             \\ \hline
PSR J2054.2+6904 & 3.7           & -             & -             & 7.5           & -             \\ \hline
PSR J2333.1-5527 & 3.1         & -             & -             & 6.9           & -             \\ \hline
\end{tabular}
\caption{Table of known Red Backs including candidates \citep{Nedreaas2024}.}
\label{tab:Red Backs}
\end{table*}

\begin{table*}
\centering
\tablenum{3}
\begin{tabular}{|c|c|c|c|c|c|}
\hline
\textbf{Name} & \textbf{dist (kpc)} & \textbf{$d_{DM}$ (kpc)} & \textbf{$\dot{E}(\times 10^{34}\, erg \, s^{-1})$} & \textbf{$P_{b}$ (h)} & \textbf{$M_{c} (M_{\odot})$} \\ \hline
B1957+20        & -             & 2.5           & 16            & 9.2           & 0.021         \\ \hline
PSR J0610-2100    & 1.5           & 3.5           & 0.85          & 6.9           & 0.025         \\ \hline
PSR J2051-0827      & 1.04          & 1.0           & 0.55          & 2.4           & 0.027         \\ \hline
PSR J0023+0923      & -             & 0.7           & 1.6           & 3.3           & 0.016         \\ \hline
PSR J0251+2606      & 1.2           & 0.96          & 1.8           & 4.85          & 0.024         \\ \hline
PSR J0636+5128    & -             & 0.5           & 0.56          & 1.6           & 0.007         \\ \hline
PSR J0952-0607      & -             & 0.97          & -             & 6.4           & 0.02          \\ \hline
PSR J1124-3653      & -             & 1.7           & 1.6           & 5.4           & 0.027         \\ \hline
PSR J1221-0633   & -             & 1.2           & 2.9           & 9.27          & 0.01          \\ \hline
PSR J1301+0833    & -             & 0.7           & 6.7           & 6.5           & 0.024         \\ \hline
PSR J1311-3430      & -             & 1.4           & 4.9           & 1.56          & 0.008         \\ \hline
PSR J1317-0157   & -             & 2.8           & 0.86          & 2.14          & 0.02          \\ \hline
PSR J1446-4701   & 1.4           & 1.5           & 3.7           & 6.7           & 0.0019        \\ \hline
PSR J1513-2550      & -             & 2.0           & 8.8           & 4.3           & 0.02          \\ \hline
PSR J1544+4937    & -             & 1.2           & 1.2           & 2.8           & 0.018         \\ \hline
PSR J1555-2908   & -             & 7.7           & -             & 5.6           & 0.05          \\ \hline
PSR J1630+3550      & -             & 1.6           & 2.64          & 7.58          & 0.0098        \\ \hline
PSR J1641+8049      & -             & 1.7           & 4.3           & 2.2           & 0.04          \\ \hline
PSR J1653-0159   & 0.84          & -             & 0.44          & 1.25          & 0.01          \\ \hline
PSR J1720-0534   & -             & 0.19          & 0.85          & 3.16          & 0.034         \\ \hline
PSR J1731-1847      & -             & 2.5           & 7.8           & 7.5           & 0.04          \\ \hline
PSR J1745+1017    & -             & 1.3           & 0.58          & 17.5          & 0.014         \\ \hline
PSR J1805+0615      & -             & 3.9           & 9.3           & 8.1           & 0.023         \\ \hline
PSR J1810+1744      & -             & 2.0           & 3.9           & 3.6           & 0.044         \\ \hline
PSR J1928+1245   & -             & 6.1           & 0.24          & 3.3           & 0.009         \\ \hline
PSR J1946-5403    & 1.23          & 0.9           & -             & 3.1           & 0.021         \\ \hline
PSR J2017-1614    & -             & 1.1           & 0.7           & 2.3           & 0.03          \\ \hline
PSR J2047+1053      & -             & 2.0           & 1.0           & 3.0           & 0.035         \\ \hline
PSR J2052+1219   & 3.9           & 2.4           & 3.4           & 2.75          & 0.033         \\ \hline
PSR J2055+3829    & -             & 4.6           & 0.36          & 3.1           & 0.023         \\ \hline
PSR J2115+5448    & -             & 3.4           & 16.5          & 3.2           & 0.02          \\ \hline
PSR J2214+3000   & -             & 3.6           & 1.9           & 10.0          & 0.014         \\ \hline
PSR J2234+0944   & -             & 1.0           & 1.7           & 10.0          & 0.015         \\ \hline
PSR J2241-5236   & 1.1           & 0.5           & 2.5           & 3.5           & 0.012         \\ \hline
PSR J2256-1024      & 2.0           & 0.6           & 5.2           & 5.1           & 0.034         \\ \hline
PSR J0336.0+7505    & -             & -             & -             & 3.7           & -             \\ \hline
PSR J1406+1222   & 1.1           & -             & -             & 1.03          & -             \\ \hline
PSR J1408.6-2917 & -             & -             & -             & 3.4           & -             \\ \hline
PSR J1838.2+3223 & 3.1           & -             & -             & 4.02          & -             \\ \hline
\end{tabular}
\caption{Table of known Black Widows including candidates \citep{Nedreaas2024}.}
\label{tab: Black widows}
\end{table*}

\newpage

\appendix

\section{Goldreich-Julian rate}

The spin-down power of a pulsar is defined as:
\begin{equation}
\dot{E(t)} = \dot{E}_{0} \left( 1 + \frac{t}{\tau_{\rm sd}} \right)^{-2}  
\label{eq:Edot}
\end{equation}
The above equation is obtained under the assumption that the energy losses of the pulsar are dominated by magnetic dipole radiation (braking index $n=3$).
In Eq.~\ref{eq:Edot}, $\dot{E_{0}}$ and $\tau_{\rm sd}$ are the initial spin-down energy and the spin-down timescale, respectively, and can be expressed as:
\begin{eqnarray}
\nonumber
\dot{E}_0 &=&\frac{8\pi^4 B_0^2 R^6}{3 c^3 P_0^4}\\
\tau_{\rm sd} &=& \frac{3 I c^3 P^2}{4\pi^2B_0^2 R^6}
\end{eqnarray}
where $P_0$ is the initial spin-down period, $B_0$ is the magnetic field at the neutron star surface while the inertial momentum is $I = 1.4\cdot 10^{45} \,{\rm g\, cm}^{2}$ and the pulsar radius $R= 10 \,{\rm km}$ \citep{Lattimer:2006xb}. 
In order to screen the electric field induced by the rotating magnetic field of the pulsars, particles are extracted from the neutron star surface at the Goldreich-Julian rate \cite{}:
\begin{equation}
\dot{N}_{\rm GJ}(t)=\frac{\sqrt{\dot{E(t)} c}}{e}
\label{eq:GJ}
\end{equation}
In this work, we focus on the study of millisecond pulsars and use typical values for this kind of object, namely $B_0 = 10^8\, G$ and $P_{0} =10^{-3} \,s$, we obtain $\tau_{\rm sd}\sim Myr$. 
We can safely neglect the time dependence and assume that $\dot{E(t)} \sim \dot{E_{0}}$.

\section{Intrabinary shock radius}
\label{IBS}
One of the possible reasons for the formation of the IBS is the interaction of the two winds.
The shock radius then can be calculated according to \cite{harding1990}. The solution of the wind pressure balance:
\begin{equation}
R_{\rm sh}=a \frac{\sqrt{A_{\rm w}}}{1+\sqrt{A_{\rm w}}}
\label{Eq:shockRadius}
\end{equation}
where $a$ is the orbital separation between the pulsar and its companion and
\begin{equation}
A_{\rm w}=\frac{\dot{E}}{\dot{M}_{\rm w} v_{\rm w} c}.
\end{equation}
is the ratio of the pulsar wind to the companion wind ram pressure.
In the above equation $v_{\rm w}$ represent the escape velocity of the companion and $\dot{M}_{\rm w}$ the mass loss rate of the companion defined in Eq.\ref{Eq: Mdot} 
The expression of Eq.~\ref{Eq:shockRadius} is valid under the assumption that $R_{sh}\ll a-R_{\rm c}$.

\end{document}